\documentclass[12pt,american]{article}


\usepackage{graphicx}
\usepackage{amsmath, amssymb, amsthm}
\usepackage{natbib}
\usepackage{subfigure}

\usepackage[mathscr]{euscript} 
\usepackage{bbm}

\usepackage{enumitem}
\setlist[enumerate]{itemsep=5pt,topsep=3pt}
\setlist[itemize]{itemsep=2pt,topsep=3pt}
\setlist[enumerate,1]{label=\arabic*.}

\renewcommand{\phi}{\varphi}
\renewcommand{\epsilon}{\varepsilon}

\usepackage{mdwlist}

\usepackage[citecolor=blue, colorlinks=true, linkcolor=blue]{hyperref}

\usepackage{caption}

\usepackage{algorithmic}
\usepackage{algorithm}

\usepackage[left=1.0in, right=1.in, top=.8in, bottom=0.8in, includehead, includefoot]{geometry}

\theoremstyle{plain}

\theoremstyle{definition}

\usepackage[multiple]{footmisc}

\usepackage{setspace}

\usepackage{authblk}

\title{Singapore's Role for ASEAN's Portfolio Investment\thanks{We would like to thank Benedict Tiu and Yupeng Wang for their excellent research assistance. We would like to thank Mariel Monica R. Sauler and Junko Koeda for helpful comments that helped us revise the paper. This research was funded by Sumitomo Mitsui Banking Corporation Foundation for International Cooperation and JSPS KAKENHI Grant Number JP23K22120. Corresponding author: Tomoo Kikuchi. Nishi-Waseda Bldg.7F, 1-21-1 Nishi-Waseda, Shinjyuku-ku, Tokyo 169-0051 Japan. Email: \texttt{tomookikuchi@waseda.jp}.}}

\author[a]{Tomoo Kikuchi}
\author[b]{Satoshi Tobe}

\affil[a]{\small Graduate School of Asia-Pacific Studies, Waseda University}
\affil[b]{\small School of Policy Studies, Kwansei Gakuin University}

\begin{document}
	
\maketitle
	
\begin{abstract} 
	
\noindent We investigate the elasticity of portfolio investment of ASEAN and OECD members to geographical distance in a gravity model utilizing a bilateral panel of 86 reporting and 241 counterparty countries/territories for 2007-2017. We find that the elasticity is more negative for ASEAN than OECD members. The difference is larger if we exclude Singapore. This indicates that Singapore's behavior is distinct from other ASEAN members. While Singapore tends to invest in distant OECD countries, other ASEAN members tend to invest in nearby countries. Our study sheds light on the role of a regional financial center in global finance.  
	
\vspace{1ex}

\noindent \textbf{Keywords:} portfolio investment; ASEAN; gravity model; Poisson Pseudo Maximum Likelihood

\noindent \textbf{JEL\ Classification:} F21; F34; O16

\end{abstract}

\clearpage

\section{Introduction}

\begin{figure}[ht!]
		\singlespacing
	\footnotesize
\begin{centering}
{\includegraphics[width=.55\textwidth]{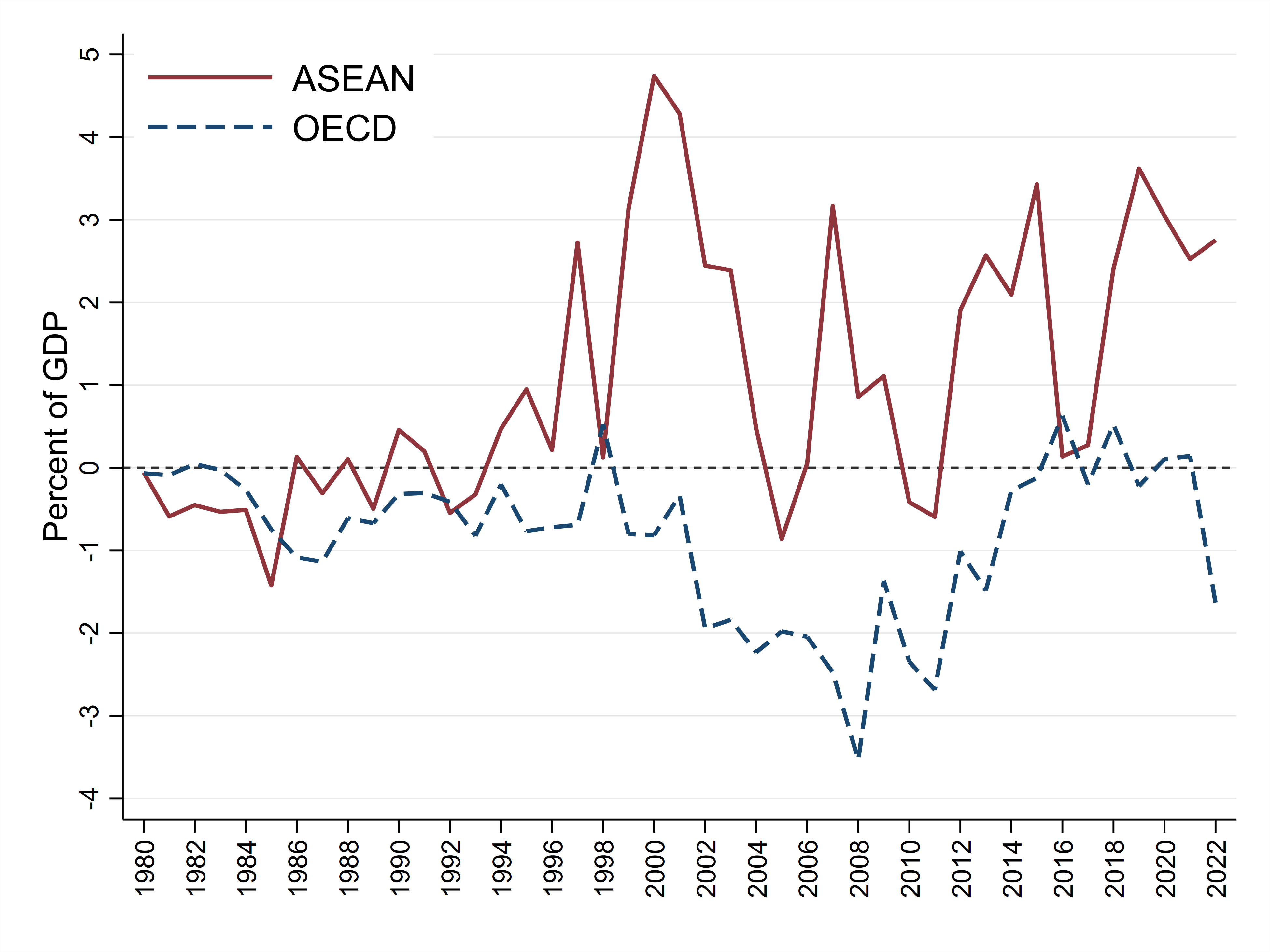}}
\caption{Net portfolio investment of ASEAN and OECD members (\% of GDP of each country group)\label{fig:FA}}		
\end{centering}	
\vspace{.5cm}
Source: IMF Balance of Payments. Note: From 1980 to 2022, ASEAN members increased from 5 to 10 and OECD members from 24 to 38.
\end{figure}

\begin{figure}[ht!]
	\singlespacing
	\footnotesize
	\begin{centering}
		\subfigure[Portfolio investment of OECD \label{fig:oecd-pi}]{\includegraphics[width=.4\textwidth]{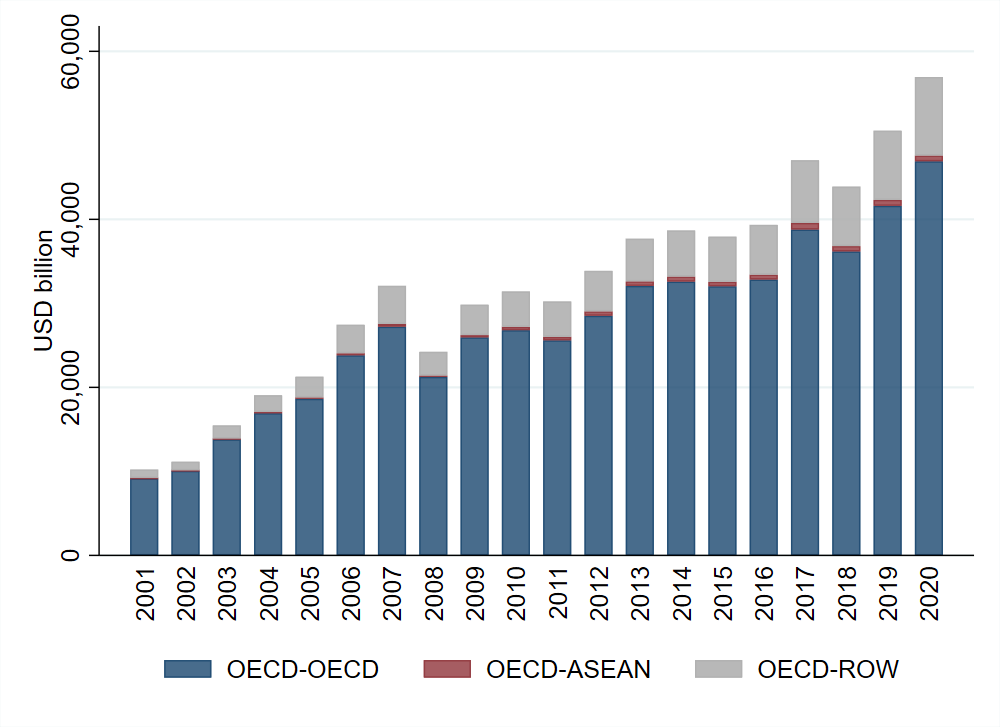}}\quad
		\subfigure[Trade of OECD  \label{fig:oecd-trade}]{\includegraphics[width=.4\textwidth]{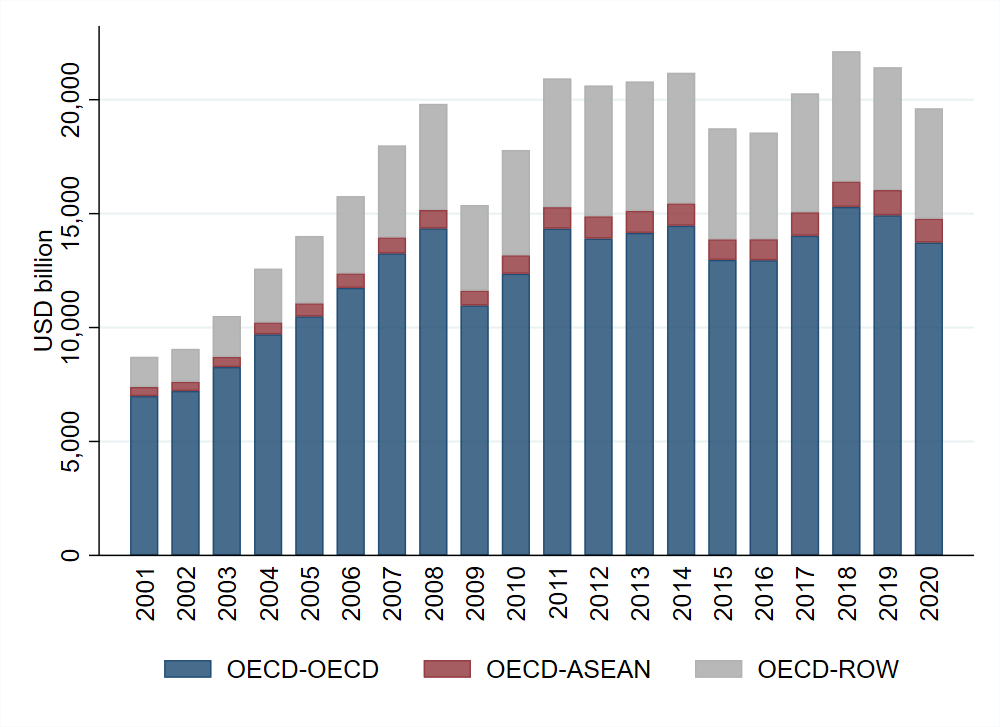}}\\
			\subfigure[Porfolio investment of ASEAN  \label{fig:asean-pi}]{\includegraphics[width=.4\textwidth]{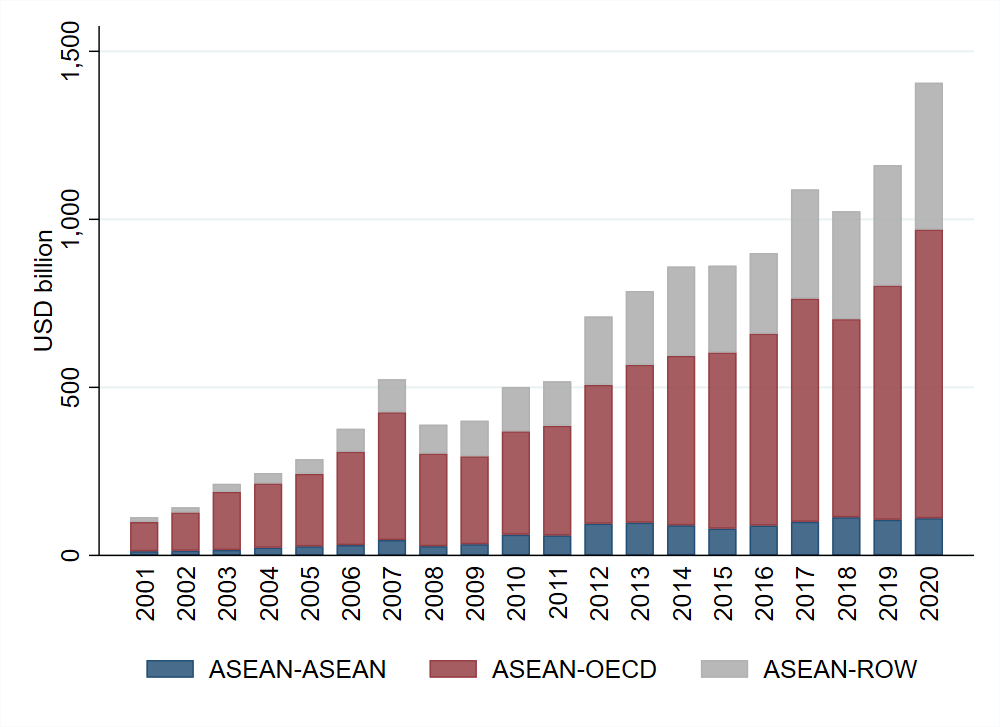}}\quad\subfigure[Trade of ASEAN \label{fig:asean-trade}]{\includegraphics[width=.4\textwidth]{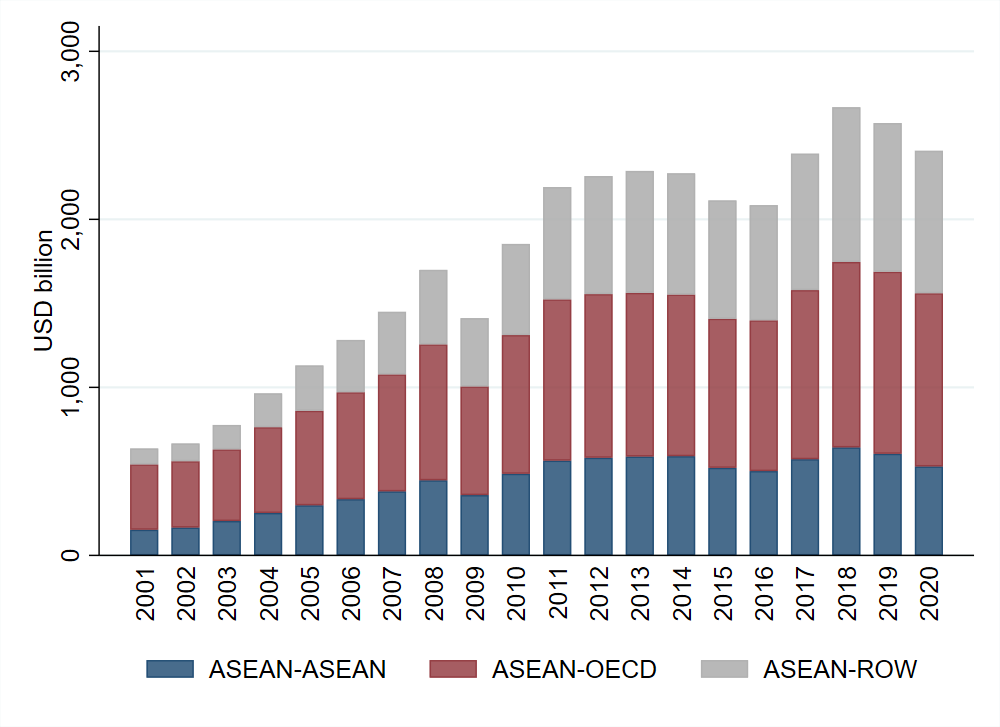}}
		\caption{Portfolio investment and trade of ASEAN and OECD members (USD billion)}
	\end{centering}	
	\vspace{.5cm}
	Source: Portfolio investment data are from IMF CPIS including 5 ASEAN countries and 34 OECD countries. Trade data are from UN Comtrade including 10 
	ASEAN countries and 34  OECD countries.	
\end{figure}

Net portfolio investment of ASEAN members started to become positive  after the Asian Financial Crisis in 1997 (see Figure \ref{fig:FA}). This means that ASEAN members invest in securities in the rest of the world more than the rest of the world invests in ASEAN members. The opposite is true for OECD members. 
In fact, OECD members hold around 80 percent of portfolio assets in other OECD members while ASEAN members hold only 7 percent of portfolio assets in other ASEAN members (60 percent in OECD members) as of 2020.
Moreover, OECD-OECD portfolio investment (from OECD members to OECD members)  has grown in tandem with OECD-OECD trade (see Figure \ref{fig:oecd-pi} and \ref{fig:oecd-trade}). On the other hand, 
ASEAN-ASEAN portfolio investment remains small relative to ASEAN-ASEAN trade (compare Figure \ref{fig:asean-pi} and \ref{fig:asean-trade}). Hence, financial market integration seems to be decoupled from good market integration for ASEAN members. 
\cite{kikuchi2021does} shows that the concentration of portfolio investment in OECD members contributes to economic growth in OECD members. Hence, it is important to understand why ASEAN members invest in securities outside more than inside the region seemingly going against the regional financial integration effort.

There seem to be gravitational forces between OECD members in capital markets while they pull portfolio investment from the rest of the world too. Since OECD members are concentrated in West Europe and North America, the average distance between OECD members is relatively small, while OECD members are relatively far from ASEAN members. 
The large portfolio investment of ASEAN members in OECD members  is puzzling as it is common to assume that information cost increases with distance even for portfolio investment \citep{martin2004, portes2005determinants, okawa2012gravity} just as transportation cost increases with distance for international trade.  

The purpose of this paper is to investigate the elasticity of portfolio investment of OECD and ASEAN members to geographical distance allowing for heterogeneity in the estimated coefficients across countries and over time. 
Unless otherwise stated, we refer to Indonesia, Malaysia, the Philippines, Singapore and Thailand, the so-called ASEAN-5, for which data are widely available as ASEAN.
We employ a gravity model approach using a bilateral portfolio investment asset dataset to examine the elasticity of portfolio investment to geographical distance. Our bilateral panel includes 86 reporting and 241 counterparty countries/territories for the period 2007-2017. 
The coordinated portfolio investment survey (CPIS) by the International Monetary Fund (IMF) reports the bilateral gross stock of 
portfolio investment (debt and equity) in each year based on the residency of investors and issuers.
We also make use of the data provided by \cite{coppola2021redrawing} who complied a restatement of the CPIS portfolio investment data from a residency to nationality basis, which is available for the period 2007-2017. 
By comparing the results of residency- and nationality-based data we highlight the role of Singapore for portfolio investment allocation of ASEAN members. 

Our main results are:
\begin{enumerate}
	\item The elasticity of both debt and equity investment to distance is more negative for ASEAN than for OECD members. This means that the inclination to invest in securities issued in nearby countries is actually stronger for  ASEAN than OECD members. The inclination is even stronger for ASEAN members excluding Singapore. This suggests that Singapore's behavior is distinct from other ASEAN members. While Singapore tends to  invest in distant countries, other ASEAN members tend to invest in nearby countries.

	\item Regarding debt investment, the elasticity  to distance has become less negative  in recent years for ASEAN members while it has become more negative for OECD members. 
	This is consistent with a dramatic increase in Singapore's debt investment in the US over the past decade (Singapore is far away from New York relative to other investment destinations). ASEAN members excluding Singapore follow a similar trend as OECD members.

	\item Regarding equity investment, the elasticity  to distance has not changed much in recent years for both ASEAN and OECD members. 
	Different coverage of countries in each data set on the residency basis suggests that China for ASEAN members and tax havens for OECD members have become significant equity investment destinations in recent years. 

\end{enumerate}

We would like to highlight two aspects of our analysis.  First, a first look at portfolio investment data shows that ASEAN members invest in distant countries compared to investment by OECD members. However, 
ASEAN members have a distance elasticity higher than OECD members. This result is obtained after controlling for country-time specific fixed effects for both reporting and counterparty countries. This means that the fixed effects, which are supposed to capture factors such as the size of GDP, the level of financial development and the institutional quality, account for why ASEAN members invest more outside than within the region. Second, our analysis highlights the role of Singapore as a platform for both inward investment from other ASEAN members and outward investment to distant OECD members. In fact, Singapore is ASEAN's largest host for multinational companies attracting portfolio investment from other ASEAN members as well as ASEAN's largest investor in the US and China. In other words, the contrasting investment behaviors of Singapore and other ASEAN members are not just caused by Singapore's investment behavior but also by Singapore being a major destination for portfolio investment of other ASEAN members and the rest of the world. This implies that ASEAN's financial integration would inevitably require a higher exposure of Singapore to securities in ASEAN members.  

Gravity model estimation is widely applied to analysis using bilateral trade flow data but also international asset allocation data such as foreign direct investment (FDI) in 
\cite{head2008fdi} and \cite{de2011does},
cross-sectional bilateral portfolio equity flows in \cite{lane2008international},
equity flows using panel data covering 1989 to 1996 in \cite{portes2005determinants}, and US bilateral asset holdings data in
\cite{chictu2014history} \citep[for a theoretical background see][]{okawa2012gravity}. Three features of our paper should be highlighted in relation to the literature. 
First, we reduce concerns about omitted variable
bias, heteroskedasticity and zero observations, which are well known challenges in estimating gravity models \citep[see][]{anderson2003gravity, silva2006log} by employing a structural gravity model estimation combining the Poissson Pseudo Maximum Likelihood (PPML) approach with
	a  set of various fixed-effects.
Second, we provide a comparison between residency and nationality-based data to account for the significance of tax havens in international financial markets in recent years.
Third, we use a comprehensive dataset covering a wide range of investor and issuer countries from 2007 to 2017 and contrast general patterns of asset allocation of ASEAN and OECD members.

The rest of the paper is organized as follows. Section \ref{sec:data} introduces the dataset. Section \ref{sec:results} introduces our baseline specification and presents our main results.  Section \ref{sec:discussons} discusses the portfolio investment of Singapore and other ASEAN members. Section \ref{sec:conlcusion} concludes.

\section{Bilateral panel data}
\label{sec:data}
We use a bilateral panel data set covering 86 reporting and 241 counterparty countries and territories from 2007 to 2017. The country lists are provided in Table \ref{fig:country_list} and \ref{fig:country_list_counterparty}.
The original bilateral portfolio investment asset data are from the CPIS
by the IMF. 
We also make use of the data provided by \cite{coppola2021redrawing} who complied a restatement of the CPIS portfolio investment data from a residency to nationality basis. 
The restatement from a residency to nationality basis is particularly important for tax havens such as the Cayman Island, Hong Kong or Singapore that attract large investment to companies that have in most cases other nationalities but reside in the tax havens. 
For example, Alibaba Group Holding Ltd. is a Chinese multinational technology company incorporated in the Cayman Islands.  When investors buy shares of the company listed in either the New York Stock Exchange (NYSE) or the Stock Exchange of Hong Kong (SEHK), it is recorded as equity investment in the Cayman Islands on a residency basis. On a nationality basis, however, the same investment is recorded as equity investment in China as its main base of operation is in China.

For China the CPIS data contain only total portfolio investment but not the decomposition into debt and equity investment. On the other hand, the 
decomposition is available for China in 
the nationality and residency-based data by \cite{coppola2021redrawing}  but the overall coverage is less comprehensive compared to the original CPIS data. For example, \cite{coppola2021redrawing}'s data miss
16 reporting countries, of which 11 are OECD countries for 2017 (see Table \ref{fig:country_list}).
Therefore, we present results using all three datasets:
1) the residency-based data  by \cite{coppola2021redrawing}, 2) the nationality-based restatement data  by \cite{coppola2021redrawing}, and 3) the residency-based original CPIS data.
The geographical distance is calculated using the latitudes and longitude of the single largest cities in countries provided by the CEPII database.

\begin{figure}[ht!]
	\begin{centering}
		\subfigure[Debt in 2007	\label{fig:bin-asean_debt2007}]{\includegraphics[width=.4\textwidth]{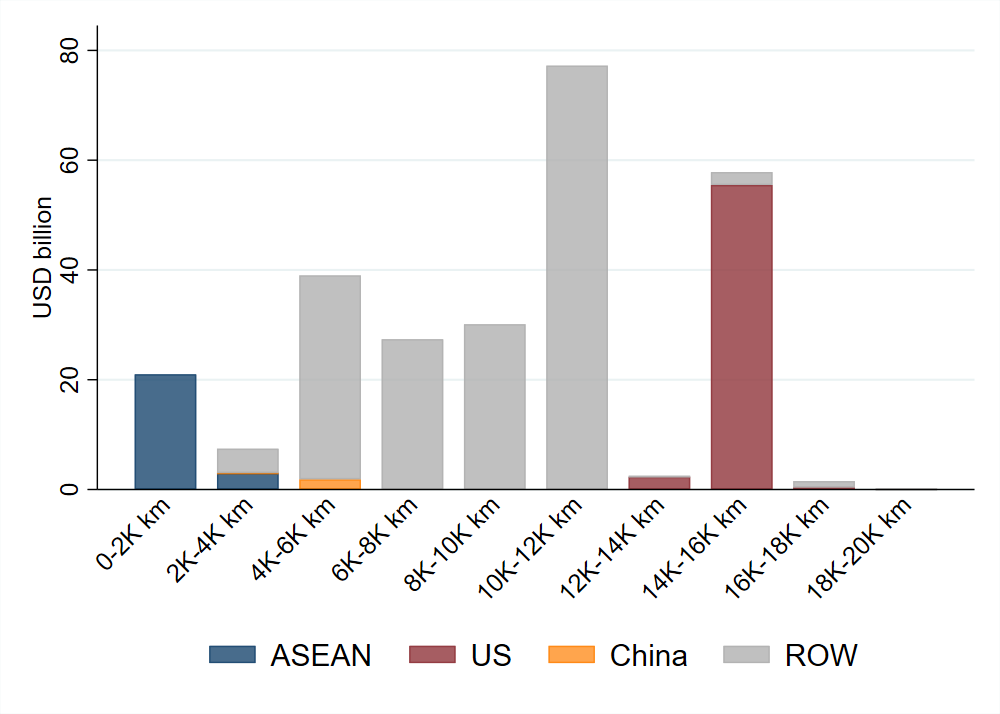}}
		\subfigure[Debt in 2017	\label{fig:bin-asean_debt2017}]{\includegraphics[width=.4\textwidth]{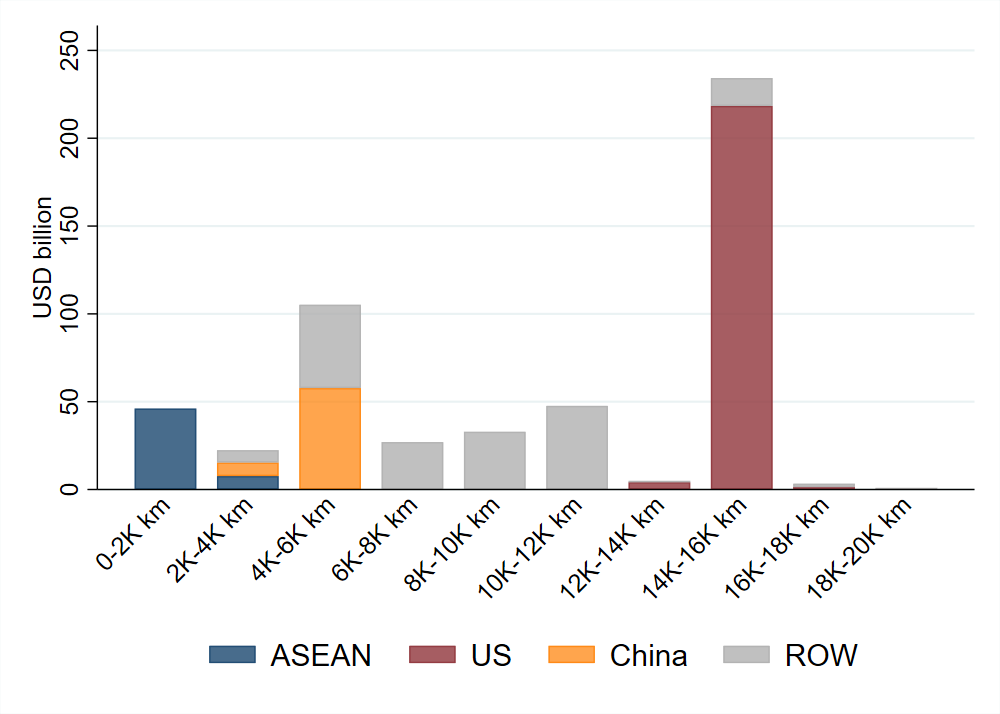}}\\
		\subfigure[Equity in 2007\label{fig:bin-asean_equity2007}]{\includegraphics[width=.4\textwidth]{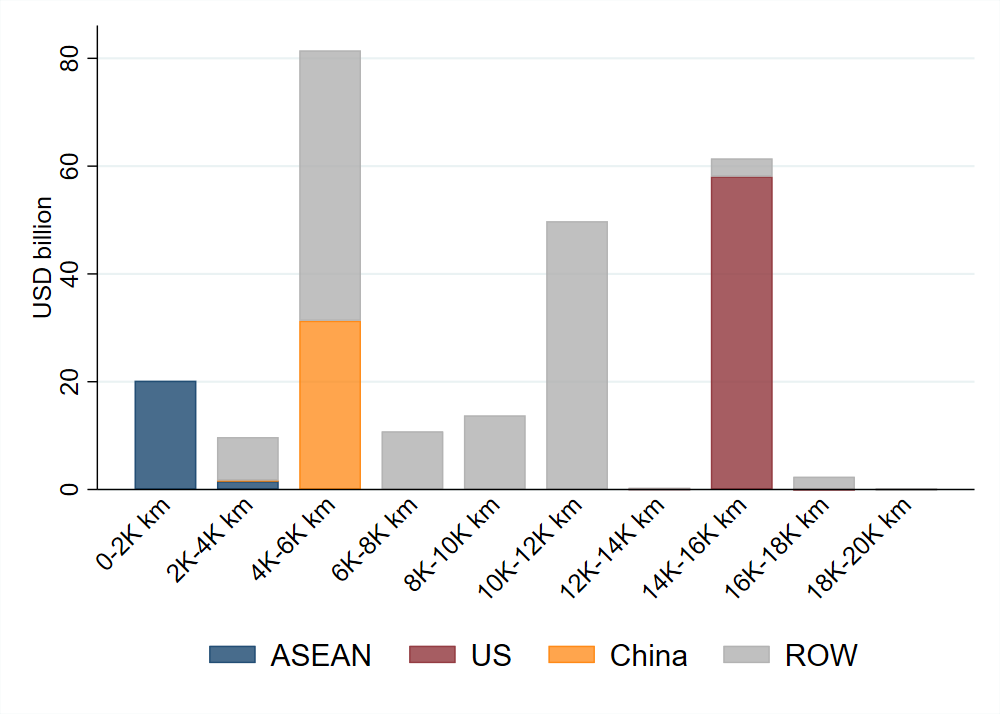}}
		\subfigure[Equity in 2017\label{fig:bin-asean_equity2017}]{\includegraphics[width=.4\textwidth]{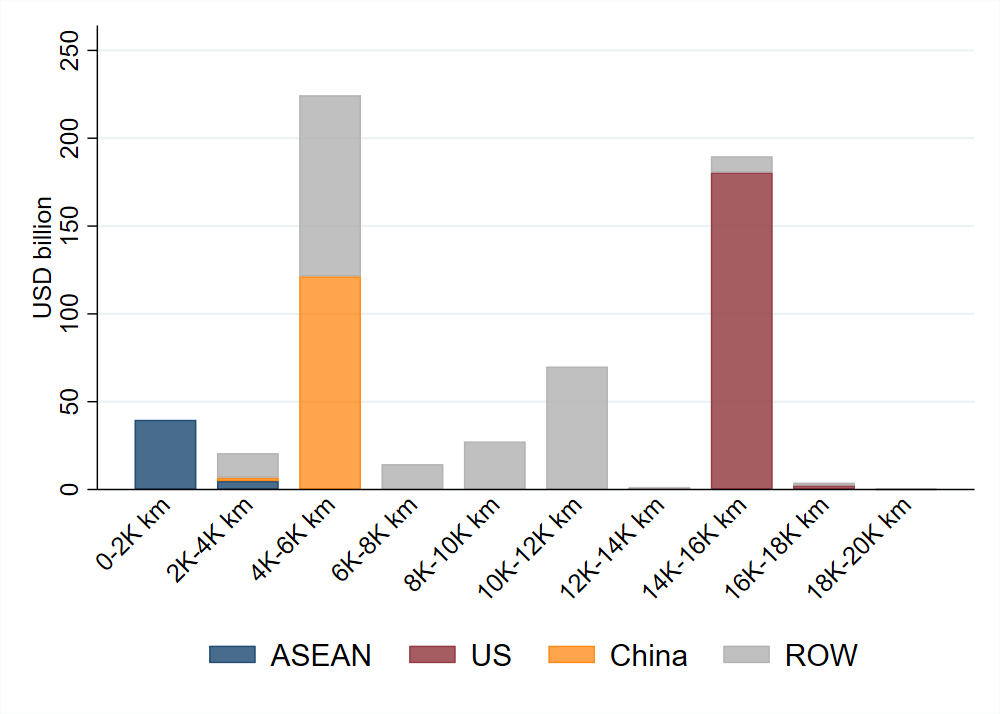}}
		\caption{Portfolio investment (nationality basis) of ASEAN members in 10 distance groups (2000km per bin, USD billion) 
		}
		\label{fig:bin-asean}
	\end{centering}	
	\vspace{.5cm}
	\footnotesize 	Source:  Distance is taken from CEPII GeoDist Database. Restated Bilateral External Portfolios - ``Tax Haven Only'' data - are based on  \cite{coppola2021redrawing} and taken from \url{www.globalcapitalallocation.com}.
	Note:  5 ASEAN source countries (Indonesia, Malaysia, Philippines, Singapore, Thailand) and 200 destination countries. 
\end{figure}

Figure \ref{fig:bin-asean} highlights the geographical asset allocation of ASEAN members.
We divide issuer countries into 10 groups based on the geographical distance from investor countries (i.e., ASEAN members). 
Each bin in the horizontal axis covers 2000 kilometers from the investors.
The first bin includes countries located from 0 to 2000 kilometers away from the investor countries, the second bin from 2000 to 4000 kilometers away and so on.\footnote{The figures are created by dividing the bilateral distance measures between the single largest cities from the CEPII GeoDist Database by 2000km in order to create a histogram with 10 bins. Counterpart countries are then divided by color into ASEAN, US, and Non-ASEAN Non-US categories.} 
The figure underscores  an inclination of ASEAN members to invest in distant countries and that the US has over the past two decades become a dominant destination for both debt and equity investments. 
In addition, the figure shows that China has become a significant equity investment destination for ASEAN members. 

\begin{figure}[ht!]
	\begin{centering}
		\subfigure[Debt in 2007\label{fig:bin-OECD_debt2007}]{\includegraphics[width=.4\textwidth]{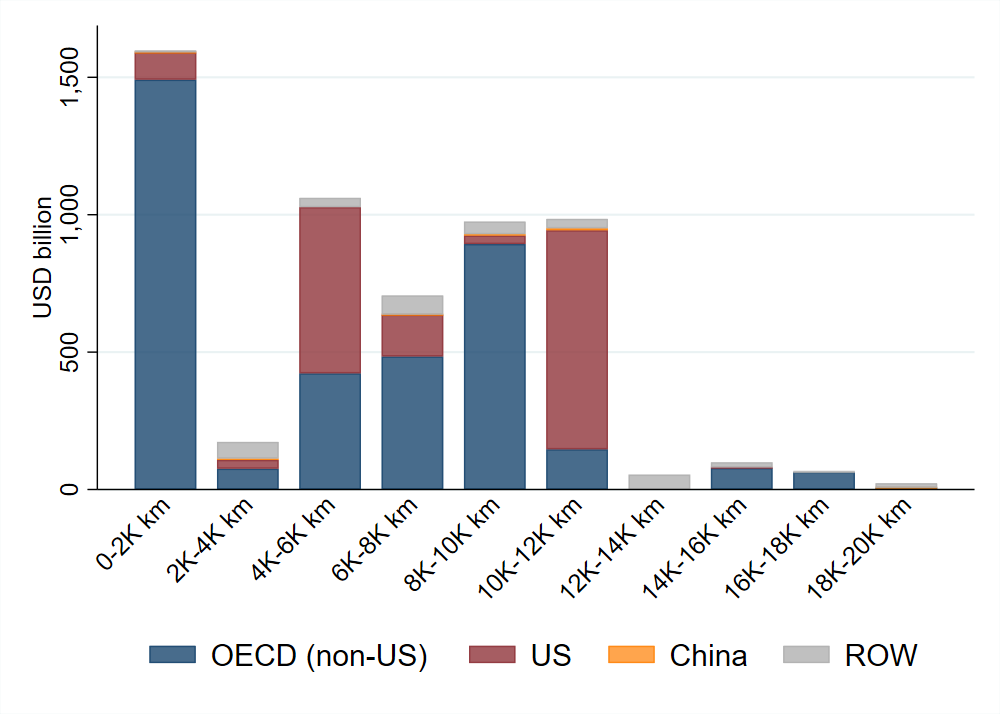}}
		\subfigure[Debt in 2017\label{fig:bin-OECD_debt2017}]{\includegraphics[width=.4\textwidth]{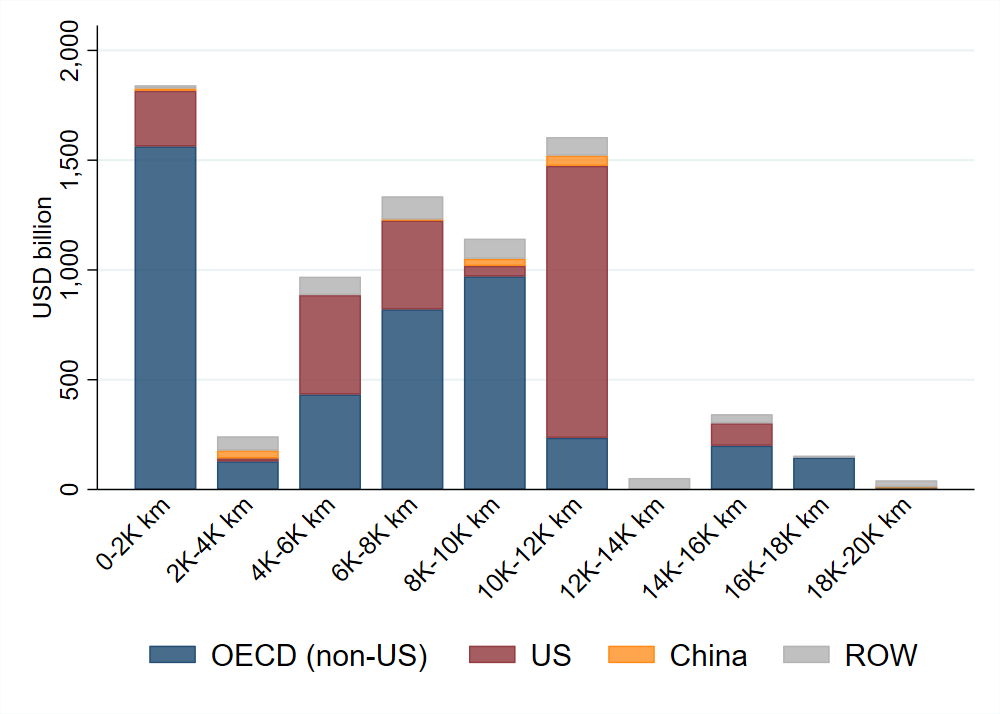}}\\
		\subfigure[Equity in 2007\label{fig:bin-OECD_equity2007}]{\includegraphics[width=.4\textwidth]{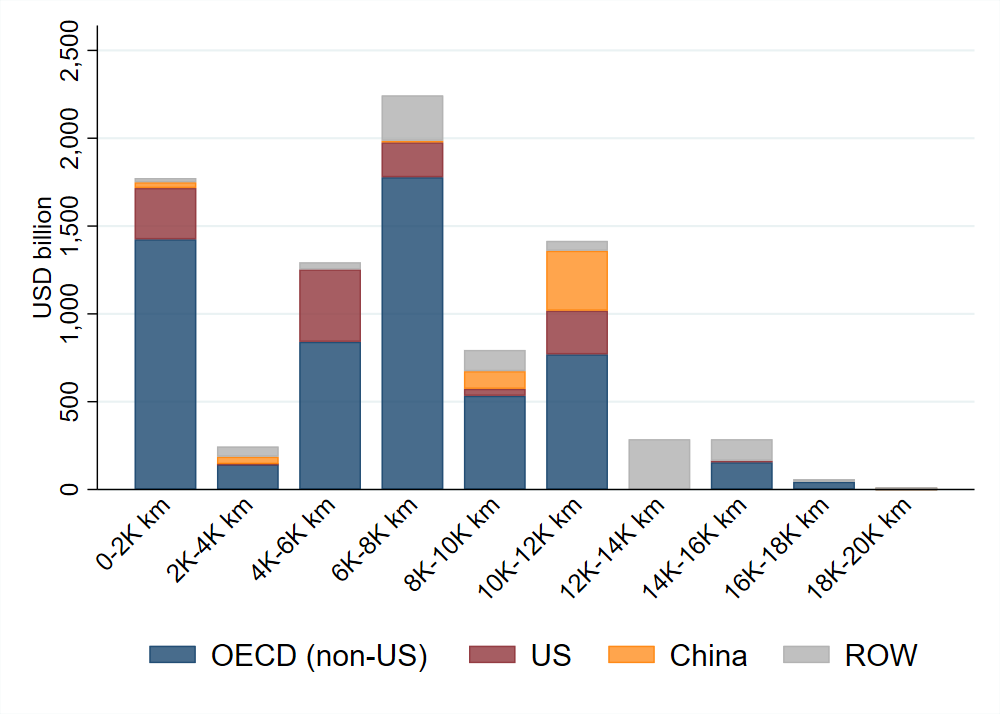}}
		\subfigure[Equity in 2017\label{fig:bin-OECD_equity2017}]{\includegraphics[width=.4\textwidth]{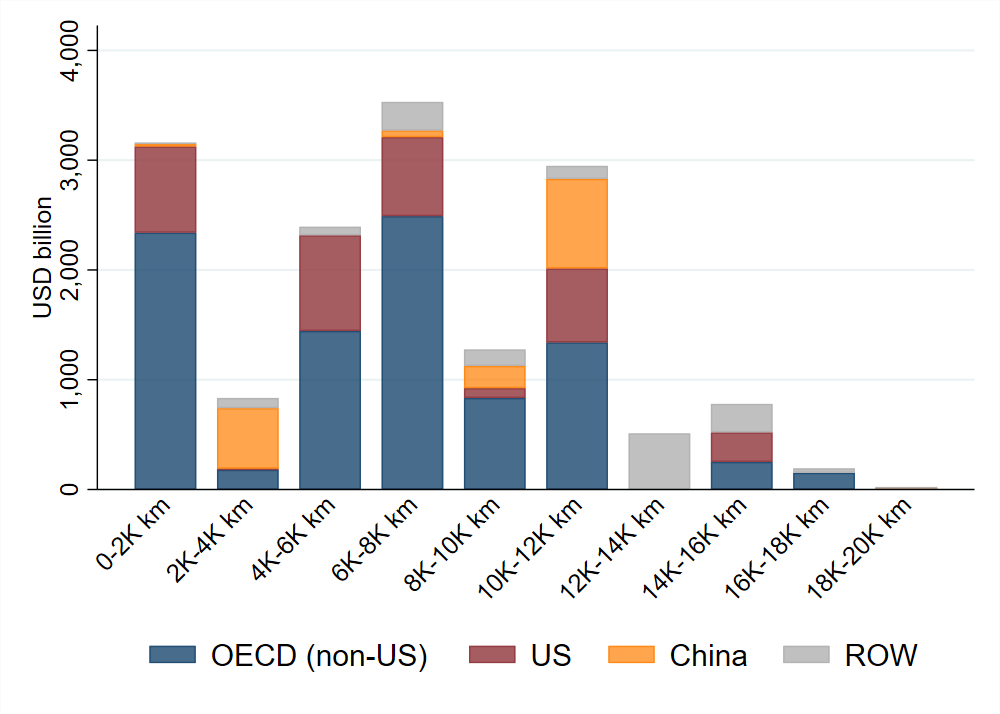}}
		\caption{Portfolio investment (nationality basis)  of OECD members in 10 distance groups (2000km per bin, USD billion) }
		\label{fig:bin-OECD}
	\end{centering}	
	
	\vspace{.5cm}
	
	\footnotesize Source: Distance is taken from CEPII GeoDist Database. Restated Bilateral External Portfolios - ``Tax Haven Only'' data - are based on \cite{coppola2021redrawing} and taken from \url{www.globalcapitalallocation.com}.
	Note:  19 OECD source countries and 200 destination countries.
\end{figure}

In contrast, OECD members have an inclination to invest in nearby countries in contrast to ASEAN members (see Figure \ref{fig:bin-OECD}). The US is a significant destination of both debt and equity investments for OECD members but unlike for ASEAN members it is in different bins for different members. In addition, we can see that China has become a significant equity investment destination for OECD members, which is also located in different bins for different members. This suggests that the inclination of ASEAN members to invest in distant countries can largely be attributed to the dominance of the US as a destination for both debt and equity investments. In addition, the significance of China as a destination for equity investment of ASEAN members should contribute to weaken the inclination of ASEAN members to invest in equity issued in distant countries.

To reduce concerns about omitted variable bias we use bilateral and unilateral dummies capturing cultural factors, colonial relationships, the legal origin of countries, and the common currency provided by the CEPII database. The common language dummy takes the value 1 if reporting and counter-party countries share common official or primary language, otherwise 0. The common language share dummy takes value 1 if the countries share a common language spoken by at least 9\% of the population, otherwise 0. 
The colonial relationship dummy takes value 1 if countries are or were in a colonial relationship post 1945, otherwise 0. The common currency dummy takes value 1 if countries use a common currency, otherwise 0. The legal system dummies capture the historical origin of a country's laws.

\section{Gravity in portfolio investment}\label{sec:results}

Section \ref{sec:data} shows that ASEAN as a whole  has a tendency to invest in distant countries while OECD has a tendency to invest in nearby countries. The result could be misleading as Singapore is by far the largest investor among ASEAN members. This section examines the country-average elasticity of portfolio investment to geographical distance for ASEAN with and without Singapore and shows that Singapore behaves differently from other ASEAN members in allocating its portfolio investment.

\subsection{Baseline analysis}

This subsection presents the baseline results.
To investigate the elasticity of portfolio investment to distance for both ASEAN and OECD  as well as the rest of the world (ROW), we estimate a gravity model using the PPML estimation method that can reduce concerns such as heteroscedasticity and zero observations. 
In particular, zero observation is a serious issue in our application because almost one half of our observations are zeros.\footnote{\cite{silva2006log} shows that the method performs well when the proportion of zeros is large by Monte Carlo simulations.}
Our baseline specification is
\begin{multline}
	P_{i,j,t}^k=\exp\left\{\beta^{ASEAN}(\ln Dis_{i,j}\times D^{ASEAN})+\beta^{OECD}(\ln Dis_{i,j}\times D^{OECD})  \right.\\[1.ex]  +                            \left.   \beta^{ROW}(\ln Dis_{i,j}\times D^{ROW}) + \beta{D}^{control}_{i,j} + \delta_{i,t}+\theta_{j,t}\right\} \epsilon_{i,j,t}, 
\end{multline}
where 
$P_{i,j,t}^k$ represents the gross stock of portfolio investment asset, and superscript $k$ corresponds to the types of portfolio
investment: debt or equity;
$Dis_{i,j}$ represents the geographical distance between the single largest cities in a particular pair country;
$D^{ASEAN}$, $D^{OECD}$ and $D^{ROW}$ are dummy variables that take a value of one if a reporting country is in ASEAN, OECD or ROW and zero otherwise; $D^{control}_{i,j}$ is a vector of bilateral or unilateral dummies capturing cultural factors, colonial colonial relationships, the legal origin of countries; $\delta_{i,t}$ and $\theta_{j,t}$ are reporting country-time specific fixed effects and counterparty country-time specific fixed effects;\footnote{We confirmed that the estimated coefficients are stable across the specifications with/without the bilateral dummies.}
$\varepsilon_{i,j,t}$ is the error term. 

Reporting and counterparty country-time fixed effects control country-specific time varying factors.
For example, they can control the sizes of GDP of investor and issuer countries in each year,
both of which are often included in traditional gravity models as well as geographical distance.
In addition to the two types of fixed effects, structural gravity models often include country-pair fixed effects \citep[e.g.][]{anderson2003gravity}.
The pair-fixed effects can control time-invariant country-pair specific factors, such as geographical distance, common official language,
contiguous borders, and presence of colonial ties between investor and issuer countries.
In this baseline analysis the specification excludes the country pair-fixed effects to focus on the average elasticity of distance throughout the sampled period.
Inclusion of both pair-fixed effects and geographical distance causes perfect collinearity, because they are time-invariant
country-pair specific variables indexed by ({i, j})-level.
Instead of the pair-fixed effects, the baseline model includes bilateral or unilateral dummy variables to reduce concerns about omitted variable bias.
We use the gross stock of portfolio investment asset just as standard gravity estimations for trade use gross export or import.
The coefficients of our interest are $\beta^{ASEAN}$, $\beta^{OECD}$, and $\beta^{ROW}$ that
capture the relative elasticity to distance of each country group.
We also estimate the alternative specification that uses ASEAN ex-Singapore dummy (i.e., ASEAN4-member dummy) instead of ASEAN5-member dummy. Each ASEAN member differs in terms of the level of economic development, the depth of domestic financial market, and the preference of investors. Especially, Singapore being an international financial center plays a special role in the group. This alternative specification will highlight the role of Singapore for portfolio investment of ASEAN.

\begin{figure}[ht!]
	\begin{centering}
		\subfigure[Debt investment\label{fig:Baseline_debt}]{\includegraphics[width=.35\textwidth]{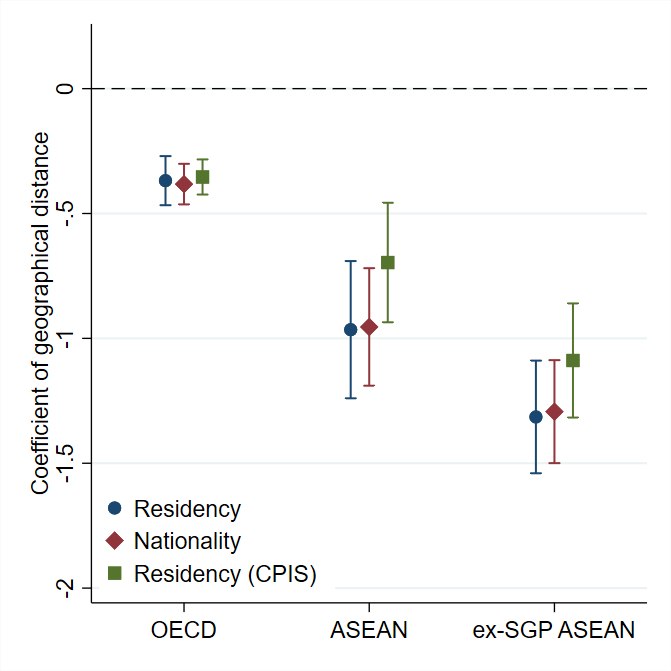}}\qquad
		\subfigure[Equity investment\label{fig:Baseline_equity}]{\includegraphics[width=.35\textwidth]{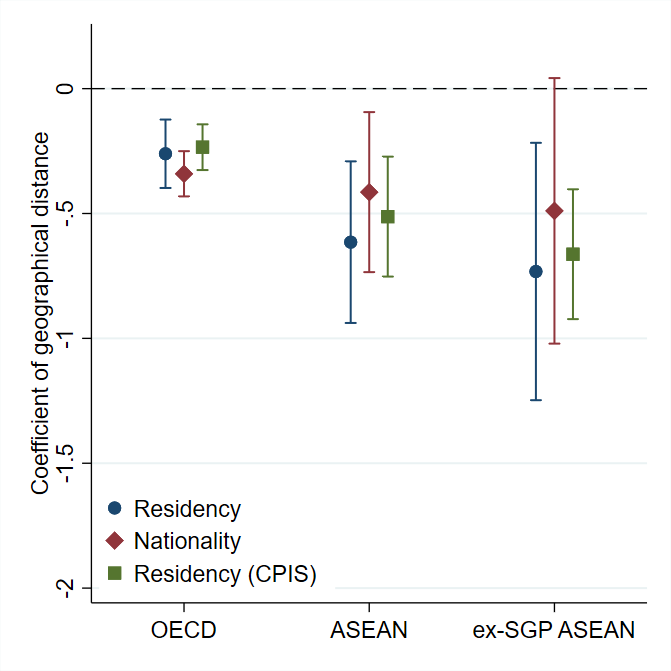}}
		\caption{Baseline analysis}
		\label{fig:Baseline}
	\end{centering}	
\vspace{.5cm}
	\footnotesize Note: The figures plot the coefficients of interaction terms of geographical distance (logged) and ASEAN/OECD dummy with 95 percent confidence intervals based on standard error clustering at country-pair level. The circle, diamond, and square markers represent the results using residency-based, nationality-based, and CPIS residency-based data. We omit the coefficients of ROW to focus on the ASEAN-OECD comparison. The full results are available on request.
\end{figure}	

Figure \ref{fig:Baseline} plots the coefficients of  distance for each country group with 95 percent confidence intervals. 
Panel (a) shows that the elasticity of distance to portfolio debt investment is negative and statistically significant
for all country groups and datasets, which is consistent with the typical behavior observed in gravity model estimations using bilateral trade flows. 
The result suggests that the investors prefer debt securities issued in a nearby country.
Moreover, ASEAN investors are more sensitive to distance than OECD investors. Panel (b) shows 
similar results for equity investment.\footnote{Table \ref{fig:baseline_app} reports the detailed results.}
These results seem to contradict our observation of ASEAN's portfolio investment in Section \ref{sec:data}. 
Therefore, Figure \ref{fig:Baseline} also presents results when the specification uses ASEAN ex-Singapore dummy (i.e., ASEAN4-member dummy)
instead of ASEAN5-member dummy.\footnote{Note that two specifications are comparable because the coefficients of OECD in the alternative specification are the same as in the baseline specification. The difference appears in the coefficients of ASEAN members and ROW, as Singapore is considered as ROW in the alternative specification.}
For debt investment the coefficients of ex-SGP ASEAN are more negative compared to
those of ASEAN.
The size of the coefficients are around -1.2 for ASEAN ex-SGP, while they are around -0.8 for ASEAN. 
Comparing the coefficients for equity investment those of  ASEAN ex-SGP are also slightly more negative than those of ASEAN (Figure \ref{fig:Baseline_equity}).
The results indicate that Singapore investors are less sensitive to distance than other ASEAN investors.\footnote{Table \ref{fig:baseline_exSGP_app} reports the detailed results.}
This distinct investment behavior of Singapore and its dominant position as both an investor and an investment destination dictate the total investment of ASEAN and explain why the investment behavior of ASEAN as a whole is different from the country-average estimates in Figure \ref{fig:Baseline}. We will further discuss the role of Singapore for portfolio investment of ASEAN in Section \ref{sec:discussons}.

\subsection{Time-series change in portfolio investment}

This subsection investigates the change in the elasticity of portfolio investment to distance over time. 
The specification is 
\begin{multline}
	P_{i,j,t}^k=\exp\left\{\sum_{t=2008}^{2017}\beta_t^{ASEAN}(\ln Dis_{i,j}\times\gamma_t\times D^{ASEAN})\right.\\  +                            \left.   \sum_{t=2008}^{2017}\beta_t^{OECD}(\ln Dis_{i,j}\times\gamma_t\times D^{OECD})\right.\\  
	+                            \left.   \sum_{t=2008}^{2017}\beta_t^{ROW}(\ln Dis_{i,j}\times\gamma_t\times D^{ROW})	+ \delta_{i,t}+\theta_{j,t} + \mu_{i,j}\right\} \epsilon_{i,j,t}.
\end{multline}
The setting follows the baseline analysis described in the previous section,
except for including the interaction terms of geographical distance ($\ln Dis_{i,j}$), the time-fixed effects ($\gamma_t$), and the ASEAN/OECD/ROW dummy ($D^{ASEAN}$, $D^{OECD}$, and $D^{ROW}$) as well as the country-pair fixed effects ($\mu_{j,i}$).
Interacting the distance and time-fixed effects enables us to include country pair-fixed effects.
The interaction terms are country-pair-time-specific variables indexed by ({i, j, t})-level,
so we can avoid multicollinearity with country-pair fixed effects indexed by ({i, j})-level.
Specification with full set of fixed effects (i.e., reporting/counterparty country-time fixed effects and country-pair fixed effects)
is the standard setting in the structural gravity literature \citep[e.g.][]{anderson2003gravity}, which can reduce the concern about possible estimation bias.
The coefficients of our interest are $\beta_t^{ASEAN}$ and $\beta_t^{OECD}$ that
capture the elasticity to distance in each year and country group
relative to a specific base year.
We set the first year of the sample, i.e., 2007, as the base year.
Thus, the sequences of  $\beta_t^{ASEAN}$ and $\beta_t^{OECD}$ capture the time variation of the elasticity 
from 2008 to 2017 in each country group.

\begin{figure}[ht!]
	\begin{centering}
		\subfigure[ASEAN \label{fig:Debt_instrument_ASEAN}]{\includegraphics[width=.45\textwidth]{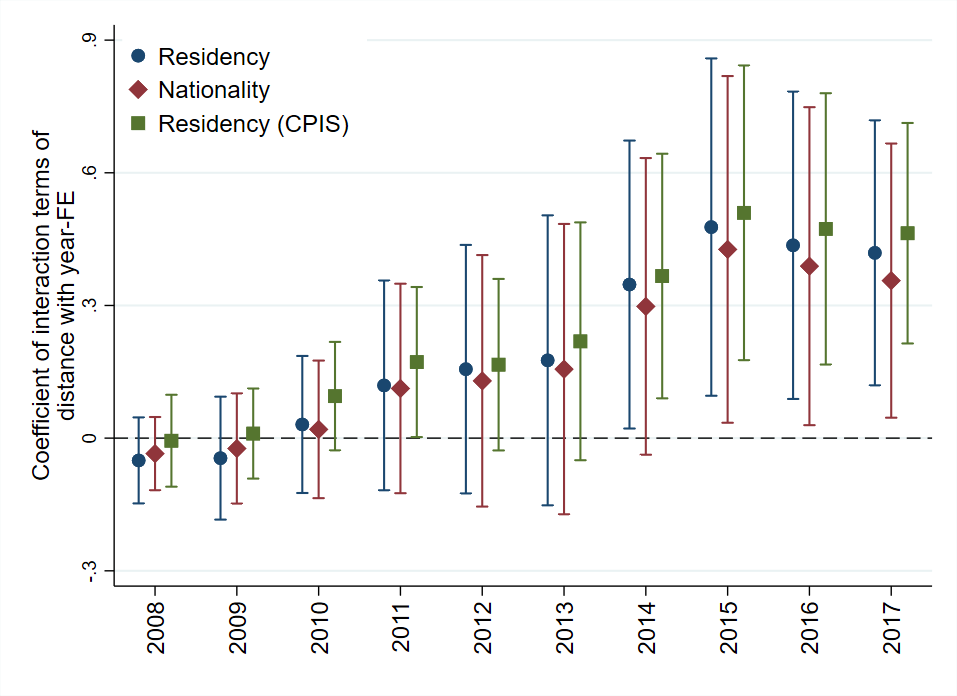}}\qquad
		\subfigure[OECD \label{fig:Debt_instrument_OECD}]{\includegraphics[width=.45\textwidth]{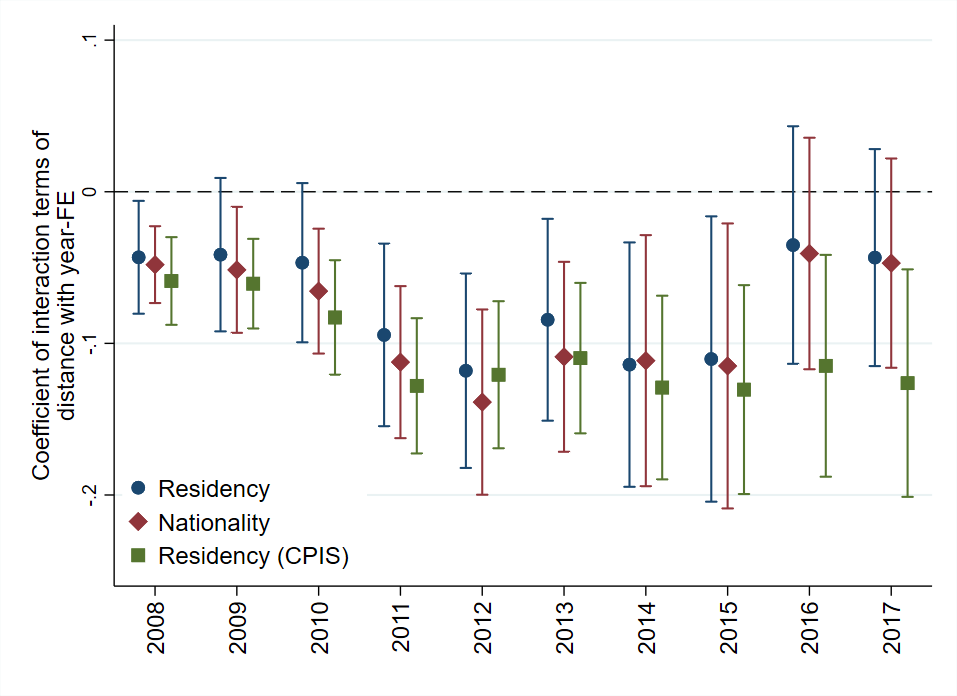}}
		\caption{Debt investment }
		\label{fig:Debt_instrument}
	\end{centering}	

\vspace{.5cm}
	\footnotesize Note: The figures plot the coefficients of interaction terms of geographical distance (logged), time-fixed effects, and ASEAN/OECD dummy with 95 percent confidence intervals based on standard error clustering at country-pair level. The circle and diamond markers represent the results using residency- and nationality-based data.
\end{figure}

Figure \ref{fig:Debt_instrument} plots  time variation of the coefficients of debt investment
to distance for ASEAN and OECD members with 95 percent confidence intervals for the residency- and nationality-based data provided by \cite{coppola2021redrawing} and the residency-based CPIS data.\footnote{Table \ref{fig:timeseries_app} reports the detailed results.}
The coefficients of ASEAN members  get larger and significant since the mid-2010,
while they are small and insignificant in the 2000s.
All three datasets follow a similar pattern.
The positive coefficients indicate that ASEAN members tend to invest in bonds issued in more distant countries than in the base year 2007. 
This is consistent with the fact that 
ASEAN members have increased portfolio investment to distant OECD countries (see Figure \ref{fig:asean-pi}) and in particular in the US in the past decade
(see Figure \ref{fig:bin-asean_debt2007} and \ref{fig:bin-asean_debt2017}).
The coefficients of OECD members present a contrasting pattern to those of ASEAN members and gets more negative and significant after the late-2000s indicating that OECD members tend to invest in bonds
issued in nearby countries more than in the base year 2007.
The three datasets largely deliver similar results.
They are consistent with the fact that OECD members have increased investment in bonds issued by other OECD members (see Figure \ref{fig:oecd-pi}) who are in relative proximity (see Figure \ref{fig:bin-OECD_debt2007} and \ref{fig:bin-OECD_debt2017}).\footnote{In the literature on bilateral trade flows, the negative coefficients of distance is known as ``distance puzzle'', where estimated
negative impact of distance on trade flows has remained persistently large across major different settings and samples (e.g.,  \citep{disdier2008puzzling, yotov2012simple}).}
	
\begin{figure}[ht!]
	\begin{centering}
		\subfigure[ASEAN\label{fig:Equity_instrument_ASEAN}]{\includegraphics[width=.45\textwidth]{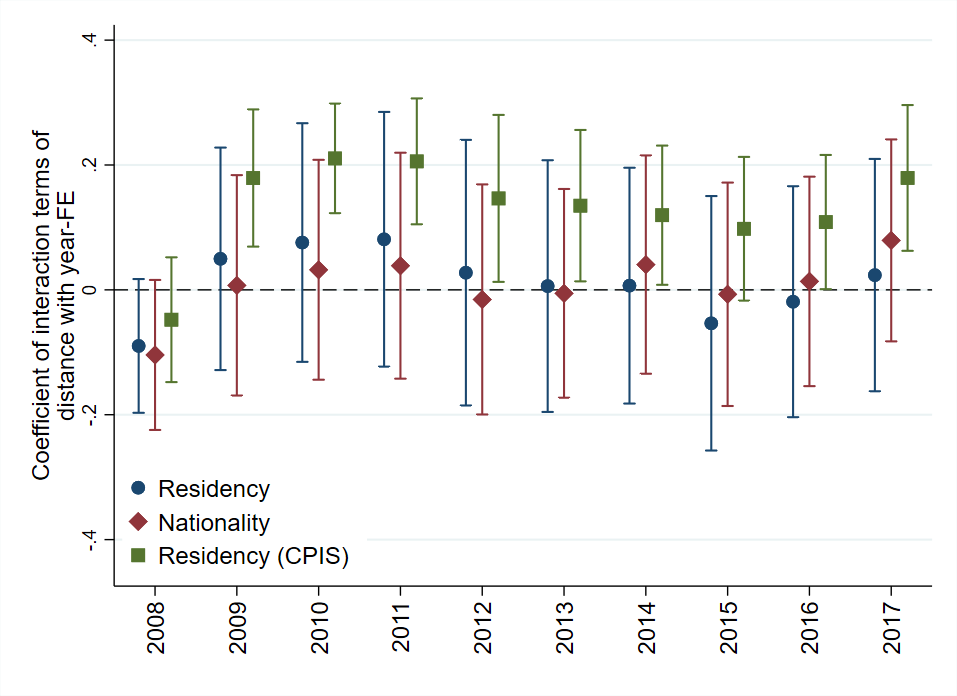}}\qquad
		\subfigure[OECD  \label{fig:Equity_instrument_OECD}]{\includegraphics[width=.45\textwidth]{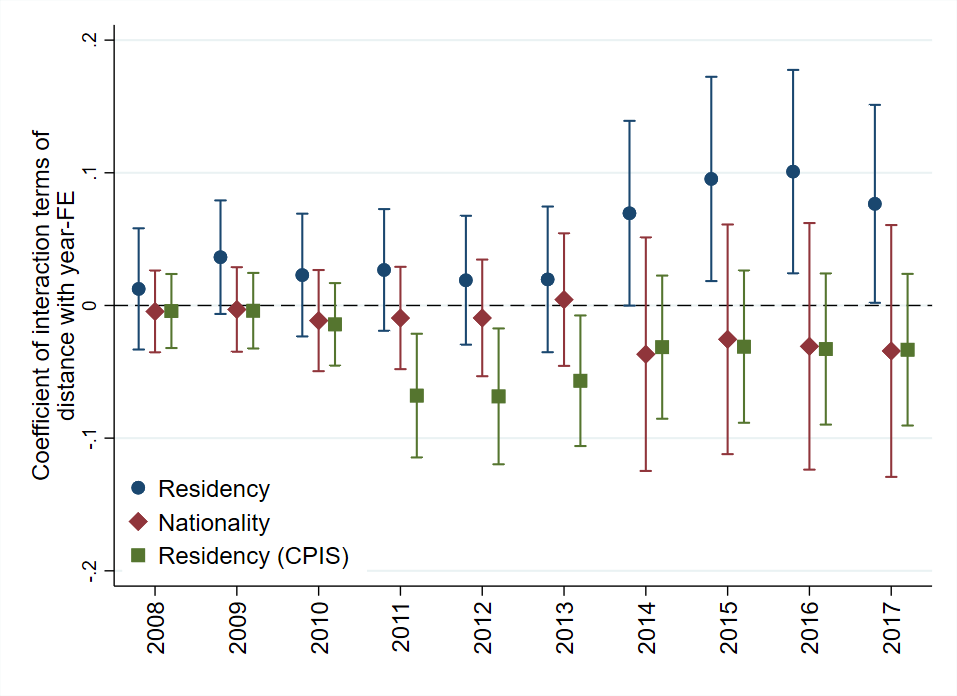}}
		\caption{Equity investment}
		\label{fig:Equity_instrument}
	\end{centering}	
\vspace{.5cm}
	\footnotesize Note: The figures plot the coefficients of interaction terms of geographical distance (logged), time-fixed effects, and ASEAN/OECD dummy with 95 percent confidence intervals based on standard error clustering at country-pair level. The circle and diamond markers represent the results using residency- and nationality-based data.
\end{figure}

Figure \ref{fig:Equity_instrument} plots the same for equity investment. We observe qualitatively similar patterns for ASEAN and OECD members especially for the nationality-based data.
The coefficients are not statistically different from zero throughout the sample period.
The results indicate little change in the geographical allocation of equity investment for ASEAN and OECD members as we observed above.
Note that the points of the residency-based data (CPIS) deviate from the other two data points based on \cite{coppola2021redrawing} in Figure \ref{fig:Equity_instrument_ASEAN} because the residency-based data (CPIS) do not provide data for equity investment in China, which is 4000-6000km away from most of ASEAN members (see Figure \ref{fig:bin-asean_equity2007} and \ref{fig:bin-asean_equity2017}).
We confirmed that the coefficients estimated by the three dataset (i.e., \cite{coppola2021redrawing}'s data on a residency-basis and a nationality-basis, and the CPIS  data on a residency-basis) are similar if we use the same sample countries.
This suggests the difference between the three dataset comes from the coverage of the countries rather than the restatement from residency- to nationality-basis.

\begin{figure}[ht!]
	\begin{centering}
		\subfigure[Debt investment\label{fig:ASEANexSGP_debt}]{\includegraphics[width=.45\textwidth]{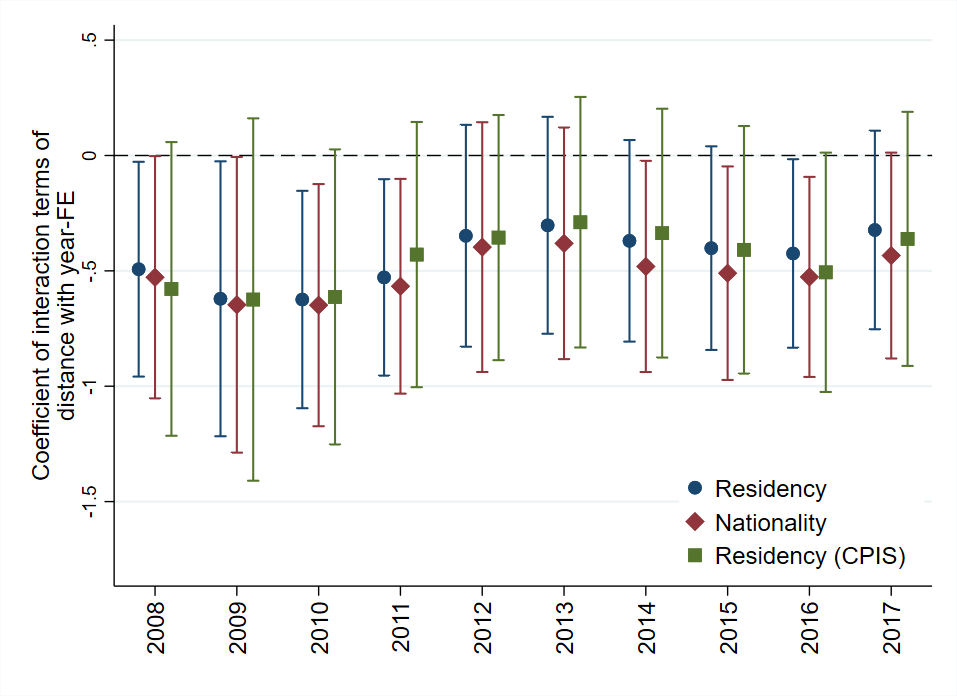}}\qquad
		\subfigure[Equity investment\label{fig:ASEANexSGP_equity}]{\includegraphics[width=.45\textwidth]{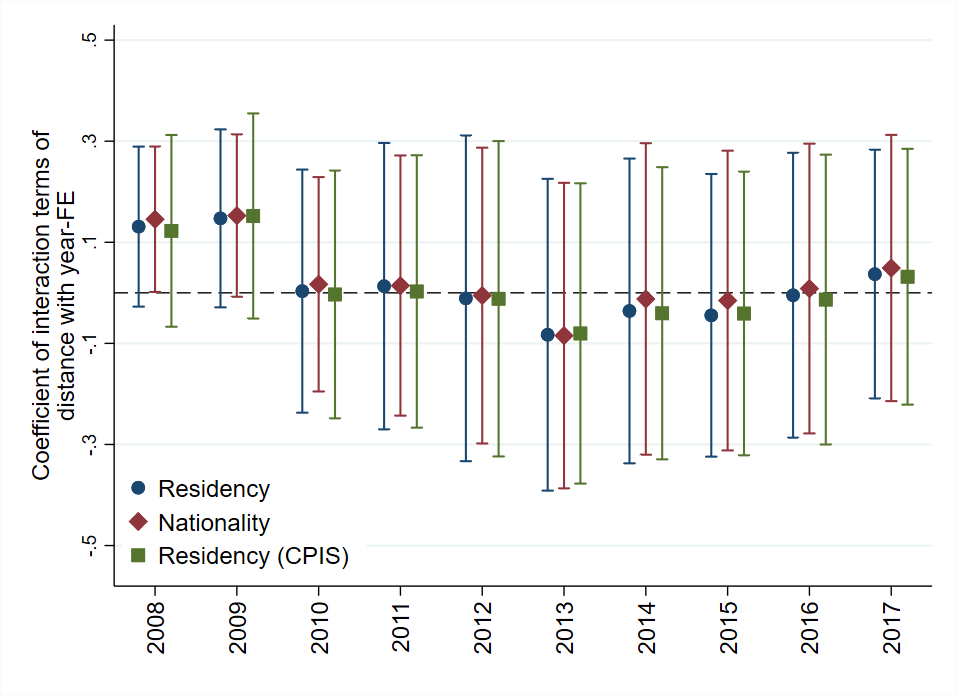}}
		\caption{Portfolio investment of ASEAN ex SGP \label{fig:ASEANexSGP}}
	\end{centering}	
\vspace{.5cm}
	\footnotesize Note: The figures plot the coefficients of interaction terms of geographical distance (logged), time-fixed effects, and ASEAN ex-Singapore dummy with 95 percent confidence intervals based on the standard error clustering at the country-pair level. The circle and diamond markers represent the results using residency- and nationality-based data.
\end{figure}	

Lastly, Figure \ref{fig:ASEANexSGP} presents results when the specification uses ASEAN ex-SGP dummy. For debt investment\footnote{Table \ref{fig:timeseries_exSGP_app} reports the detailed results.}
the coefficients get negative and weakly significant after the late-2000s, which is similar to the result for OECD members presented in
Figure \ref{fig:Debt_instrument_OECD}.
The result for equity investment is similar to that for OECD members too.
These results indicate that investors in ASEAN members except Singapore 
have not significantly changed their behavior 
in allocating portfolio investment across countries since the base year 2007, especially for equity investment.
Therefore, the positive trend in the elasticity of ASEAN's debt investment to distance  observed in Figure \ref{fig:Debt_instrument_ASEAN} must be driven by Singapore.

\section{Discussions}
\label{sec:discussons}

Section \ref{sec:results} shows that the elasticity of portfolio investment to distance is more negative for ASEAN than OECD. However, the difference is less significant when we exclude Singapore from the ASEAN sample. This suggests that Singapore's investment behavior is distinct from other ASEAN members. Therefore, this section examines the role of Singapore for portfolio investment of ASEAN.

\subsection{Singapore as a global investor}
\label{subsec:foreign_bias}

\begin{figure}[ht!]
	\begin{centering}
		\subfigure[2007	\label{fig:network_asean_countries_equity2001}]{\includegraphics[width=.3\textwidth]{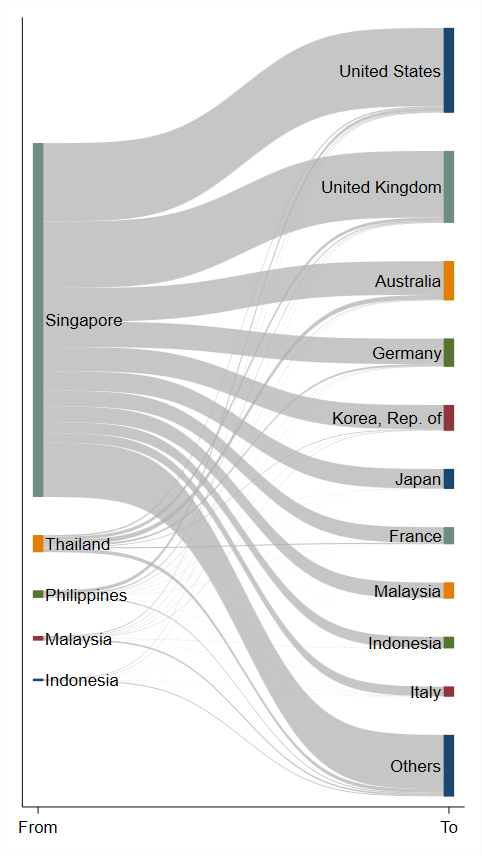}}\qquad
		\subfigure[2017	\label{fig:network_asean_countries_equity2020}]{\includegraphics[width=.3\textwidth]{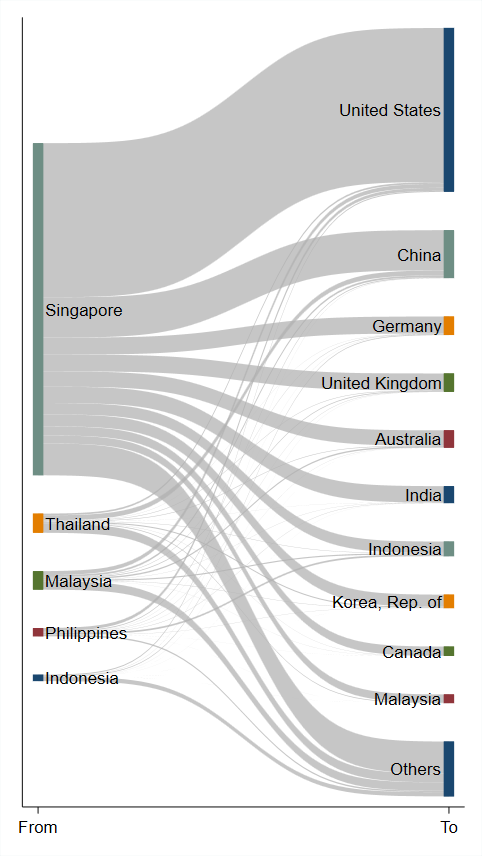}}
		\caption{Debt investment (nationality basis) of ASEAN members in top 10 countries }
		\label{fig:network_asean_countries_debt}
	\end{centering}	
\vspace{.5cm}
	\footnotesize Note: The size of each ``thread'' denotes a relative size for each graph (year), so we can not compare the investment size across graphs (years). Source: Restated Bilateral External Portfolios - ``Tax Haven Only'' data based on  \cite{coppola2021redrawing} and obtained from \url{www.globalcapitalallocation.com}. 
	
\end{figure}

Singapore is a global debt investor allocating 85 percent of its debt investment to OECD members in 2007 and 72 percent in 2017.
Singapore is also by far ASEAN's largest investor in foreign debt and the US grew to become the dominant destination of its debt investment from 2007 to 2017 (see Figure \ref{fig:network_asean_countries_debt}). Indeed, the US alone accounts for 43 percent of ASEAN's debt investment in 2017.\footnote{Malaysia and Indonesia feature within ASEAN's top 10 debt investment destinations. However, the amount is dwarfed by Singapore's investment in the US and China.} Notably, emerging as the number two destination  China accounts for 13 percent of ASEAN's debt investment in 2017.  
This largely explains the positive trend in the elasticity of ASEAN's debt investment  shown in Figure \ref{fig:Debt_instrument_ASEAN}.

\begin{figure}[ht!]
	\begin{centering}
		\subfigure[2007\label{fig:network_asean_countries_equity2001}]{\includegraphics[width=.3\textwidth]{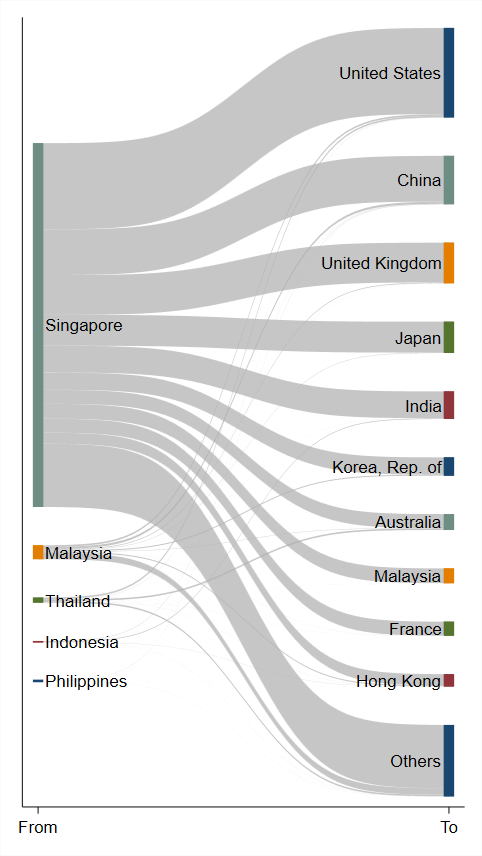}}\qquad
		\subfigure[2017\label{fig:network_asean_countries_equity2020}]{\includegraphics[width=.3\textwidth]{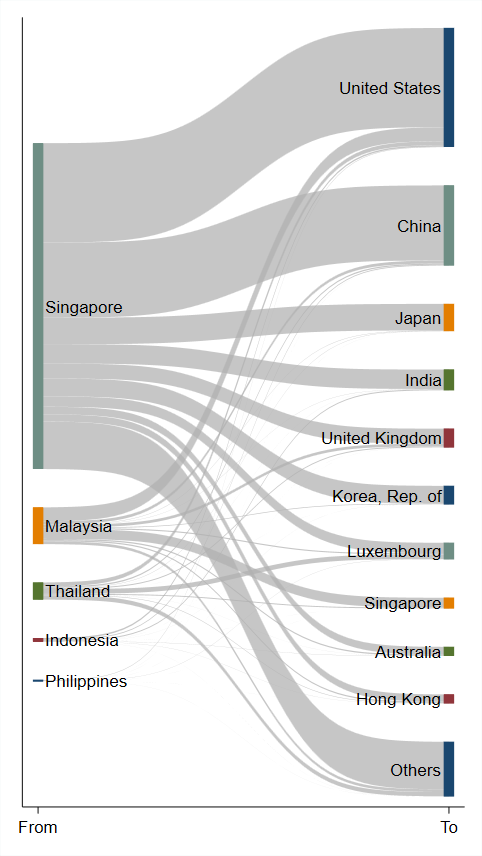}}
		\caption{Equity investment (nationality basis) of ASEAN countries in top 10 countries}
		\label{fig:network_asean_countries_equity}
	\end{centering}	
	\vspace{.5cm}
	\footnotesize Note: The size of each ``thread'' denotes a relative size for each graph (year), so we can not compare the investment size across graphs (years). Source: 200 destination countries. Restated Bilateral External Portfolios - ``Tax Haven Only'' - data based on \cite{coppola2021redrawing} and obtained from \url{www.globalcapitalallocation.com}.
\end{figure}	
Singapore is also by far the largest equity investor among ASEAN members. 
The US and China have grown to become the most dominant destinations accounting for 31 and 21 percent of ASEAN's equity investment in 2017, most of which is Singapore's investment (see Figure \ref{fig:network_asean_countries_equity}). Note that the US is 14000-16000km away  and China is 4000-6000km away from most ASEAN members. The presence of China as a destination for Singapore's equity investment and its relative proximity to Singapore explain why the gap in the value of coefficients between ASEAN and OECD is not as large for equity as for debt investment  (as shown in Figure \ref{fig:Baseline}). Turning to intra-regional investment, Malaysia was Singapore's  largest equity investment destination in ASEAN in 2007, but it has since then fallen out of top 10 equity investment destinations of ASEAN members. On the other hand, Singapore features as a top 10 destination of ASEAN's equity investment in 2017.

\subsection{Multinationals in Singapore}\label{sec:singapore}

\begin{figure}[ht!]
	\begin{centering}
		\subfigure[Debt\label{fig:stack-asean-debt}]{	{\includegraphics[width=.4\textwidth]{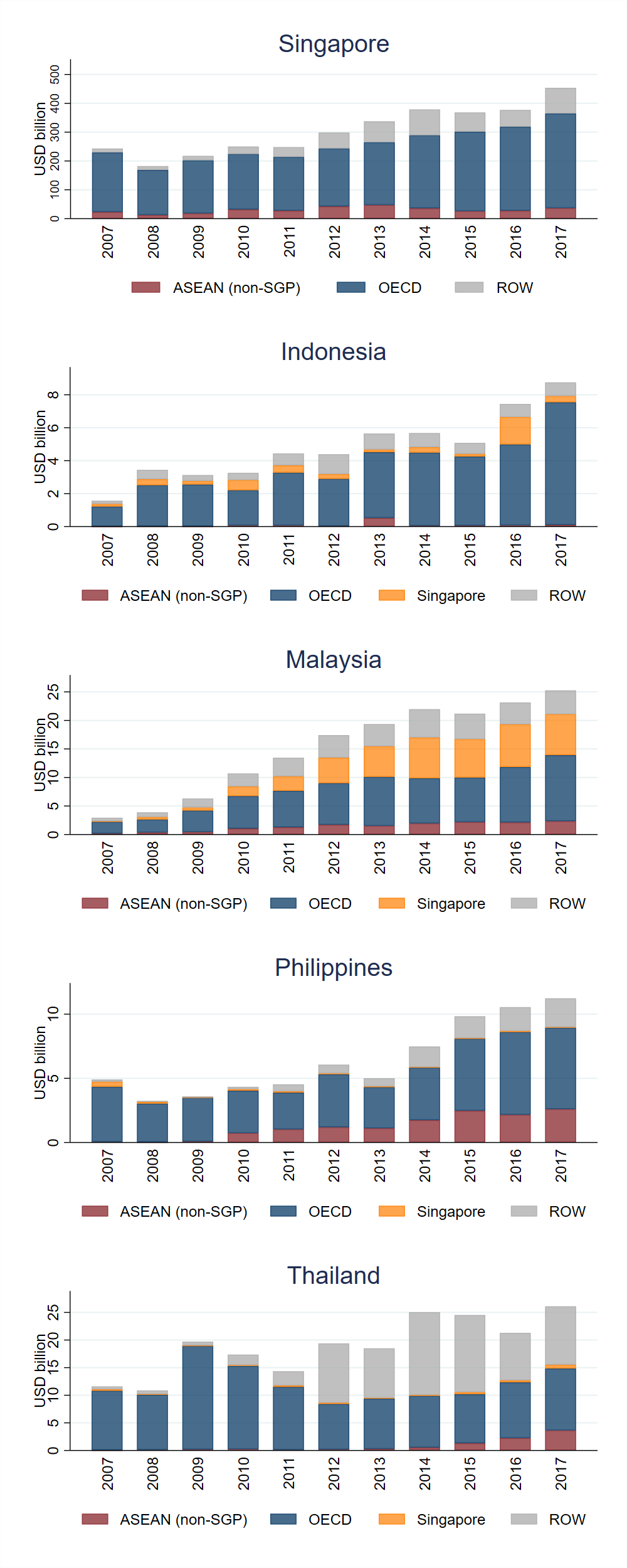}}
		}
		\subfigure[Equity\label{fig:stack-asean-equity}]{	{\includegraphics[width=.4\textwidth]{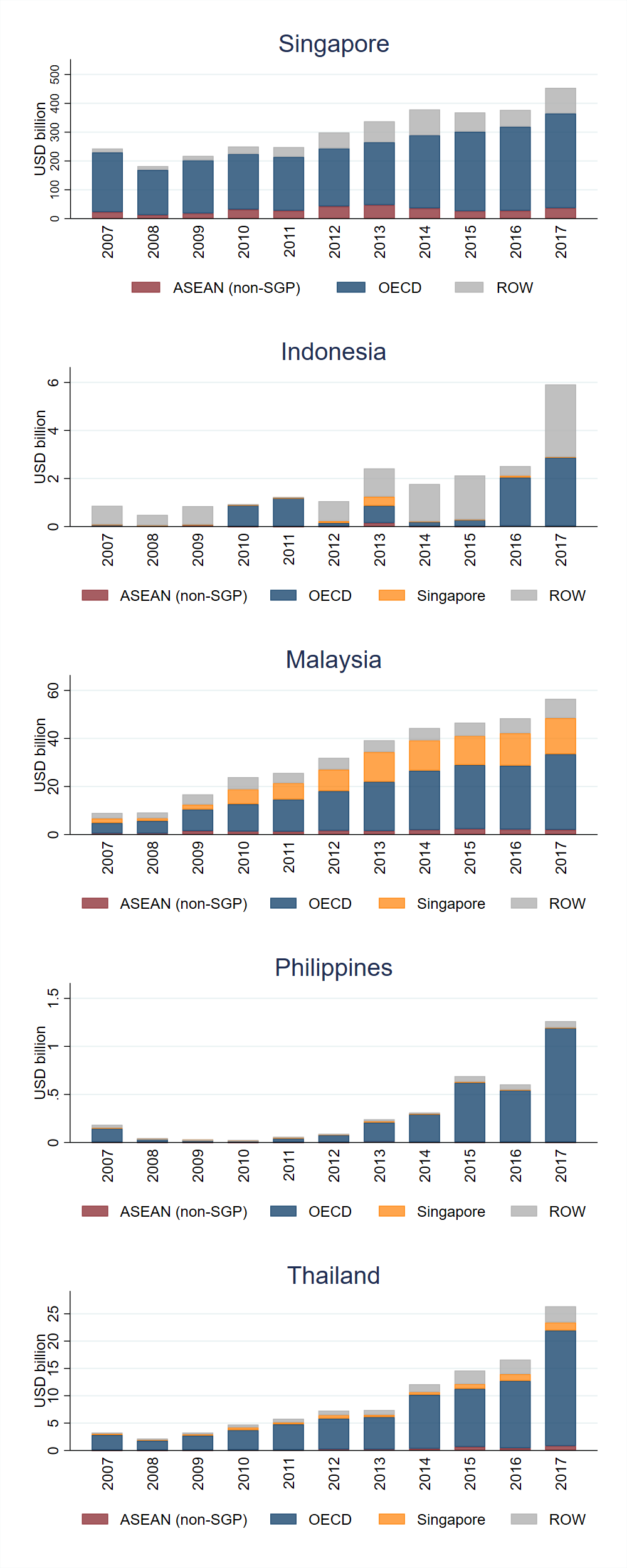}}
		}	
		\caption{Portfolio investment (nationality basis) of ASEAN countries from 2007 to 2017 (USD billion) \label{fig:stack-asean}}
	\end{centering}	
	\vspace{.5cm}
	\footnotesize Note: 200 destination countries. Restated Bilateral External Portfolios - ``Tax Haven Only'' data - based on the work by \cite{coppola2021redrawing} and obtained from \url{www.globalcapitalallocation.com}.
\end{figure}

Considering its size of GDP relative to  other ASEAN members, Singapore's portfolio investment is disproportionately large as shown in Section \ref{subsec:foreign_bias}. 
On the other hand, Singapore is also a major destination of ASEAN's portfolio investment. Time series of portfolio investment of ASEAN members on a nationality basis from 2007 to 2017 show that for Malaysia Singapore is a major destination for both debt and equity investment (see Figure \ref{fig:stack-asean}). 
Given Malaysia's relatively large investment size 
we can say that most of ASEAN's portfolio investment in ASEAN members goes to Singapore.

\clearpage

We now examine the changes the restatement (i.e., from a residency to nationality basis) makes to ASEAN's portfolio investment. The change is calculated by subtracting the residency-based data from the nationality-based data.
The changes highlight  1) ASEAN's investment in multinational companies residing in Singapore and 2) Singapore's investment in multinational companies in other tax havens such as the Cayman Island.\footnote{The database provided by \cite{coppola2021redrawing}  identifies the identity of the issuer of debt or equity but not of the investor. Therefore, it can not tell us the nationality of multinational companies who reside in Singapore and invest outside Singapore.}

\begin{figure}[ht!]
	\begin{centering}
		\subfigure[Investment from ASEAN \label{fig:res-nat-top10-source2}]{\includegraphics[width=.4\textwidth]{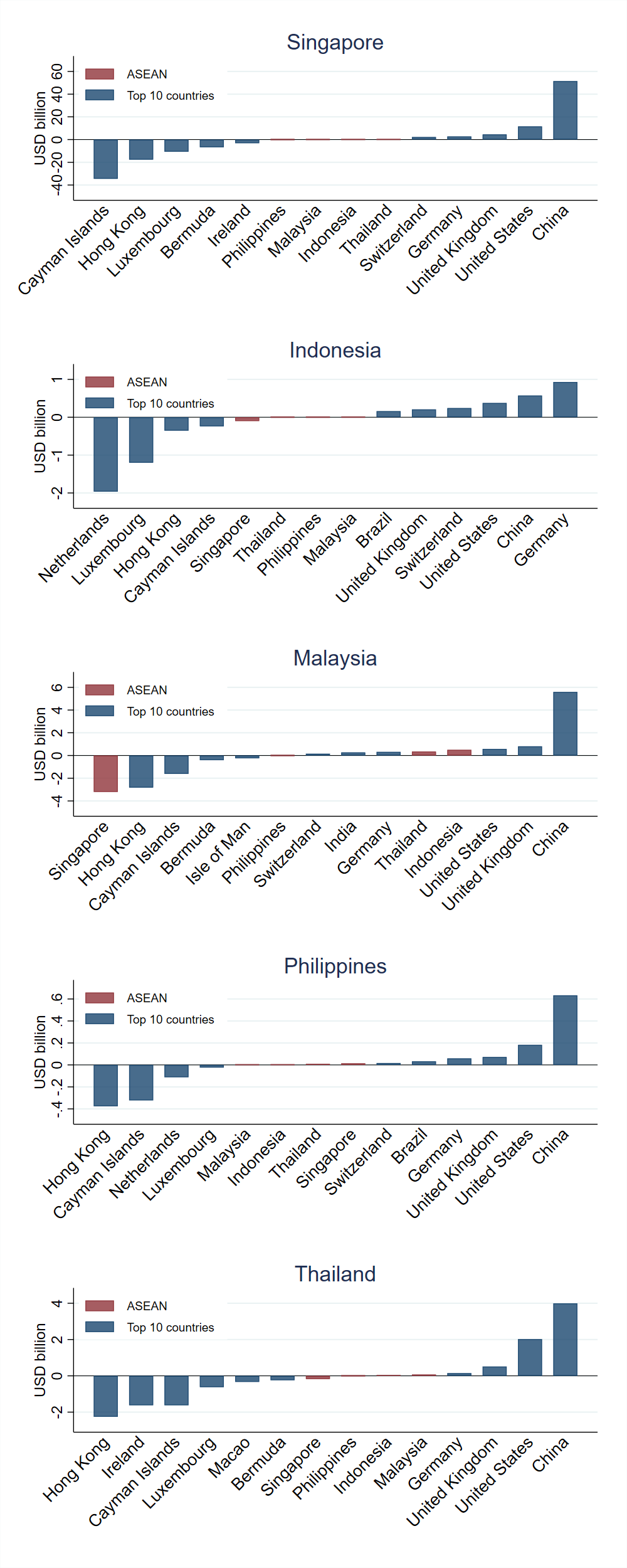}}\qquad
		\subfigure[Investment into ASEAN   	\label{fig:res-nat-top10-destination2}]{\includegraphics[width=.4\textwidth]{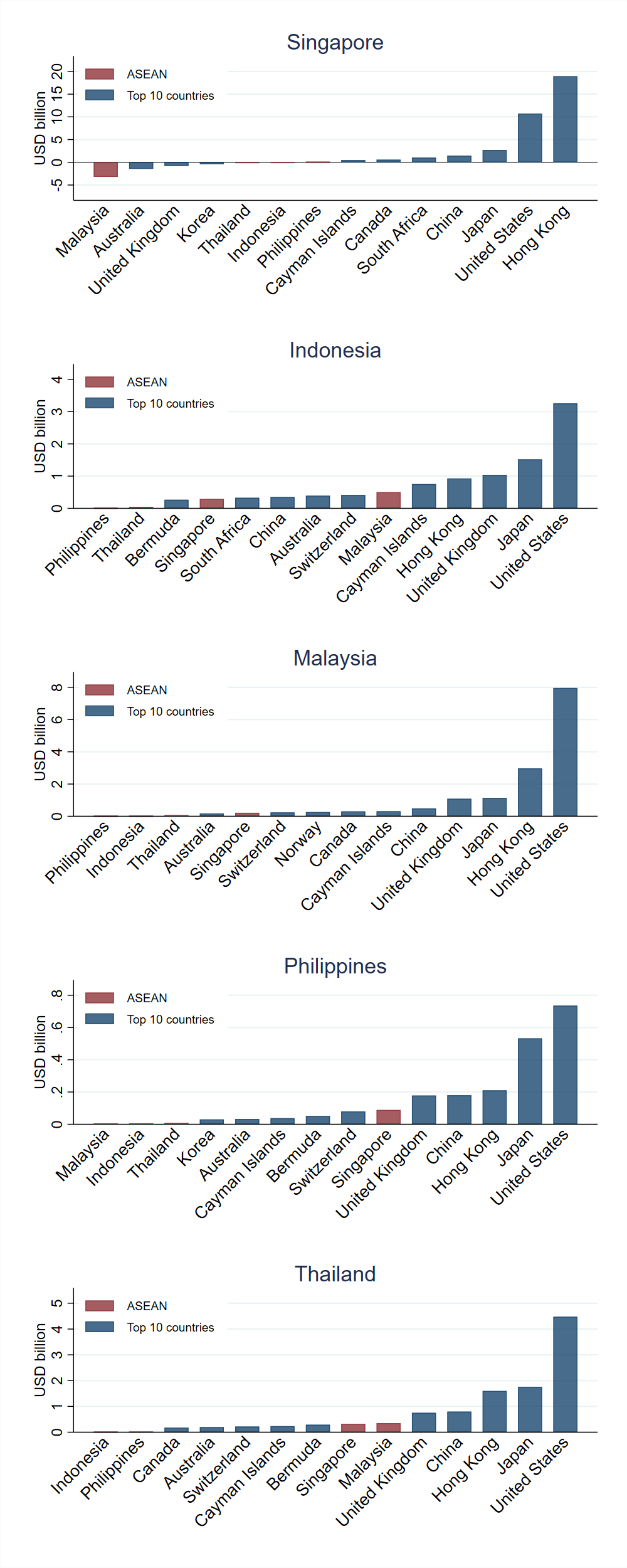}}
		
		\caption{Portfolio investment from and into ASEAN: Top 10 changes (nationality - residency) in 2017 (USD million)}
		
	\end{centering}	
	\vspace{.5cm}
	\footnotesize Note: 200 destination countries. Restated Bilateral External Portfolios - ``Tax Haven Only'' data based on the work by \cite{coppola2021redrawing} and obtained from \url{www.globalcapitalallocation.com}.
\end{figure}

Singapore records by far the largest changes but for all ASEAN countries the restatement increases investment in China and the US and decreases investment in tax-haves such as the Cayman Island, Singapore, Hong Kong, and the Netherlands (see Figure \ref{fig:res-nat-top10-source2}). This suggests a significant amount of investment from ASEAN in Chinese and American companies located in tax-havens  such as Singapore's investment in Chinese companies in the Cayman Island and Hong Kong (e.g.~Alibaba Group Holding Ltd. and Tencent Holdings Ltd.). We estimate that roughly 50 billion USD out of Singapore's portfolio investment of 70 billion USD on a residency basis in the Cayman Islands and Hong Kong are in Chinese companies.\footnote{If we assume that the entire drop in Singapore's investment in Hong Kong (20 billion USD) and in the Cayman Islands (30 billion USD) is due to investment in Chinese companies, we get roughly 50 billion USD.}

\begin{figure}[ht!]
	\begin{centering}
		{\includegraphics[width=.6\textwidth]{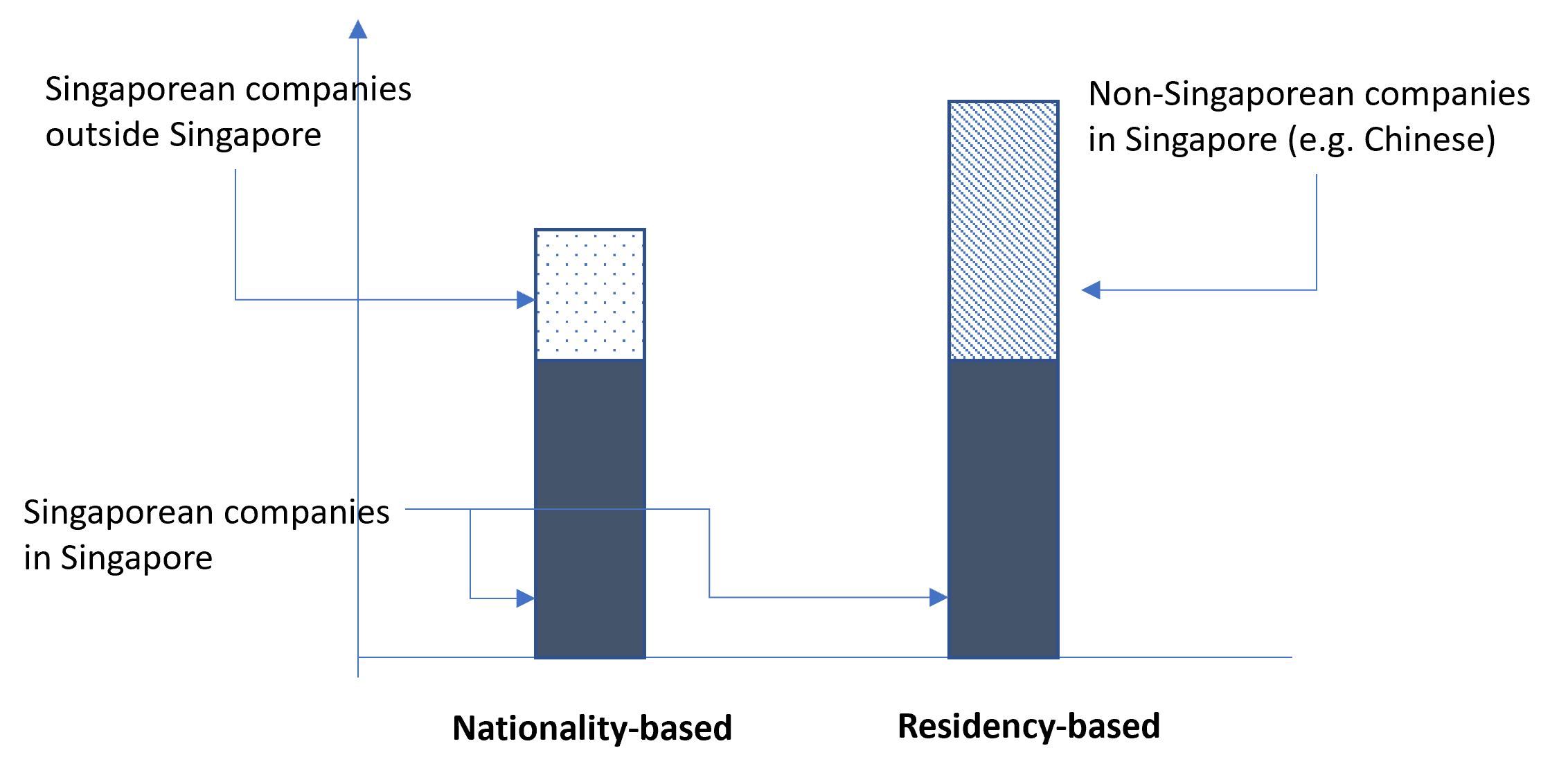}}
		\caption{Malaysia's investment in Singapore}
		\label{fig:restatement-example}
	\end{centering}	
\end{figure}

Next to Singapore, Malaysia records the largest changes. Malaysia's investment  after the restatement falls the largest in Singapore (see Malaysia in Figure \ref{fig:res-nat-top10-source2}). 
This shows that Singapore is the largest host of Malaysia's investment in multinational companies outside their home countries.
On the other hand, Malaysia's investment  after the restatement rises the largest in China (see Malaysia in Figure \ref{fig:res-nat-top10-source2}) suggesting that Malaysia invests a significant amount in Chinese companies residing in Singapore (see Figure \ref{fig:restatement-example}). We estimate that roughly 2 billion USD out of Malaysia's total portfolio investment of 25 billion USD on a residency basis in Singapore in 2017 are in Chinese companies.\footnote{If we assume that Malaysia invests in Chinese companies in Hong Kong, the Caymand Islands and Singapore and that most of the drop in Malaysia's investment in Hong Kong (2.5 billion USD) and two thirds of the drop in the Cayman Islands (1 billion USD) is due to investment in Chinese companies, we get roughly 2 billion USD ($5.5-2.5-1=2$).}

Turning to investment into ASEAN members after the restatement in Figure \ref{fig:res-nat-top10-destination2} Singapore records the largest changes but the difference to other ASEAN members is not as large as in Figure \ref{fig:res-nat-top10-source2}. The largest rise in investment in all ASEAN members after the restatement is from the US, Hong Kong, and Japan showing that they  invest in ASEAN companies outside ASEAN more than in multinational companies in ASEAN.
Notably, Malaysia records the largest fall for investment in Singapore (see Singapore in Figure \ref{fig:res-nat-top10-destination2}) showing that Malaysia is the largest international investor in such multinational companies in Singapore.    
The example shows that Singapore is a platform for multinational (e.g.~Chinese)  companies to raise capital attracting inward portfolio investment from other ASEAN members.

To summarize, Singapore's investments in tax havens such as the Cayman Island and Hong Kong, which are largely investments in Chinese and American companies, are by an order of a magnitude larger than other ASEAN member's investments in those tax havens. 
On the other hand, ASEAN members (Malaysia in particular) invest in multinational (Chinese in particular) companies residing in Singapore. 
Note that those investment in multinational (non-Singaporean) companies in Singapore as well as Singapore's investments in multinational companies in tax havens  highlight Singapore's role as a platform for both inward and outward investments in multinational companies (whose nationality is different from residency) but are only a part of Singapore's overall outward investments as well as investments of other ASEAN members in Singapore.

\section{Conclusion}
\label{sec:conlcusion}

It is often argued why ASEAN members tend to invest in securities outside more than inside the region.  We estimate a gravity model using the PPML estimation method utilizing a bilateral panel of 86 reporting and 241 counterparty countries/territories for the period 2007-2017. We find that the elasticity of both debt and equity investments to geographical distance is actually more negative for ASEAN than for OECD members after controlling for country-time fixed effects. 
However, we find that the ASEAN-OECD difference in elasticity gets smaller when we exclude Singapore from the ASEAN sample.  This indicates that Singapore's investment behavior is distinct from that of other ASEAN members. Singapore tends to invest in distant countries such as the US while other ASEAN members tend to invest in nearby countries. 
Since Singapore's investment is disproportionately large, it drives the total investment of ASEAN members. There may be foreign investors who choose Singapore as a global investment platform. Unfortunately, the available datasets do not allow us to identify the nationality of investors. In addition to facilitating 
outward investment to OECD members, Singapore also receives
inward investment from ASEAN members. The restatement data allow us to better understand Singapore's role as a financial center. Multinational companies in Singapore attract investment from ASEAN members, in particular, Malaysia.  On the other hand, Singapore is by far ASEAN's largest investor  in American and Chinese companies residing in tax haves such as the Cayman Island and Hong Kong.
Lastly, our analysis shows that the gap in the elasticity between ASEAN and OECD members is smaller for equity than for debt investment reflecting the increasing presence of China as a destination for equity investment by ASEAN members.

\clearpage

\appendix
\counterwithin{figure}{section}
\counterwithin{table}{section}

\section{Appendix}
\label{sec:appendix}

\begin{table}[ht!]
	\begin{centering}
		{\includegraphics[width=1.0\textwidth]{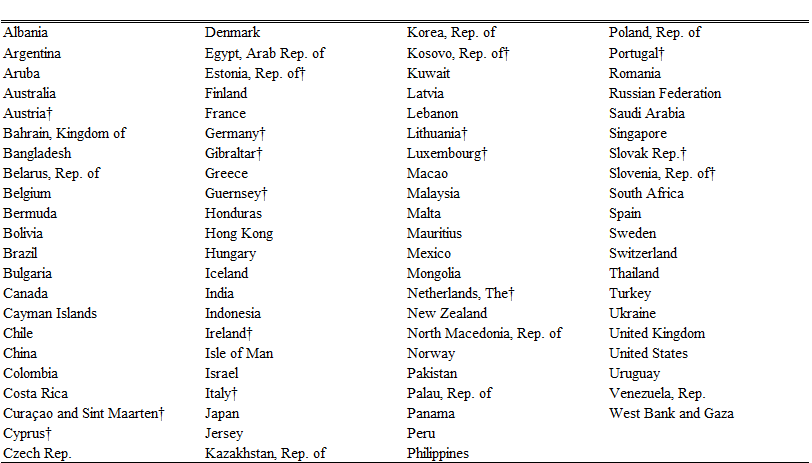}}
		\caption{List of reporting countries}
		\label{fig:country_list}
	\end{centering}	
	\footnotesize Note: Data availability of reporting country as of 2017. {$\dagger$}Reporting countries \cite{coppola2021redrawing}'s data do not cover.
\end{table}	

\begin{table}[ht!]
	\begin{centering}
		{\includegraphics[width=1.0\textwidth]{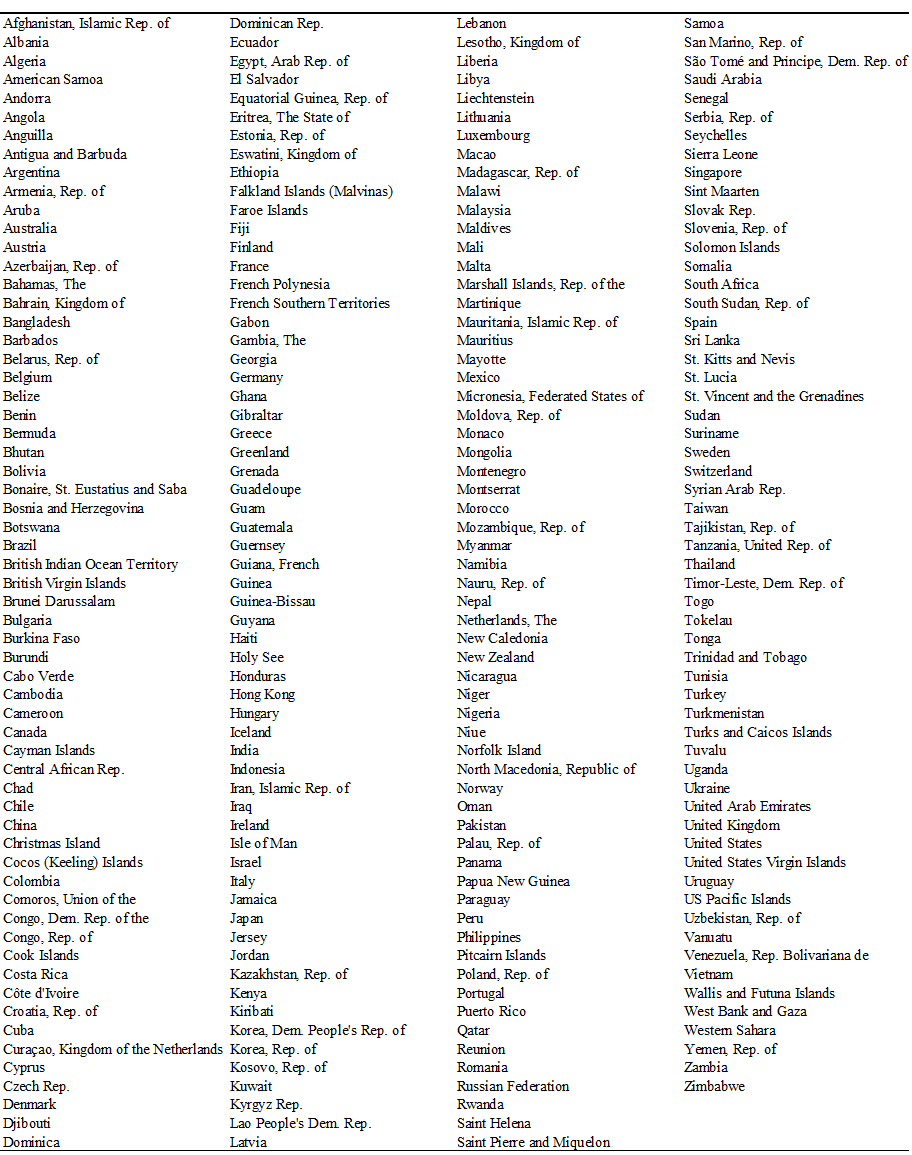}}
		\caption{List of counterparty countries}
		\label{fig:country_list_counterparty}
	\end{centering}	
	\footnotesize Note: Data availability of counterparty countries as of 2017. 
\end{table}	

\begin{table}[ht!]
	\begin{centering}
		{\includegraphics[width=0.8\textwidth]{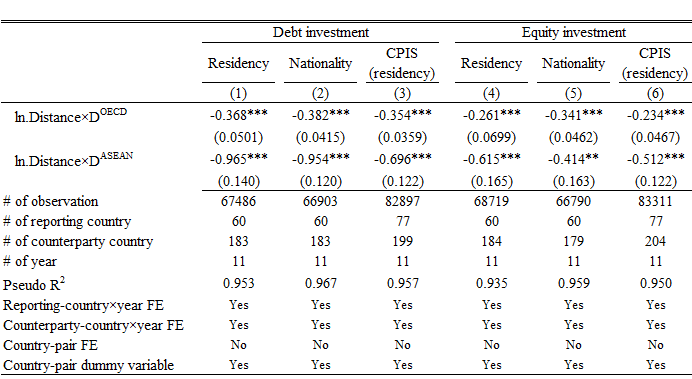}}
		\caption{Baseline analysis}
		\label{fig:baseline_app}
	\end{centering}	
	\footnotesize Note. Standard errors in parenthesis are clustered at county-pair level.  ***, ** and * denote signifinance at the 1, 5 and 10 percent levels. Constants and the coefficients for the rest of the world are omitted.
\end{table}	

\begin{table}[ht!]
	\begin{centering}
		{\includegraphics[width=0.8\textwidth]{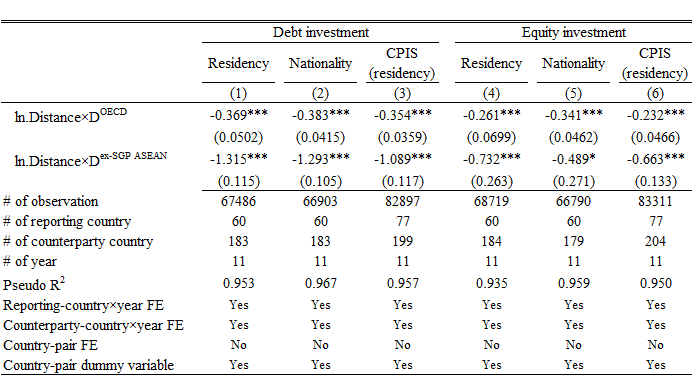}}
		\caption{Baseline analysis (ASEAN ex-SGP dummy)}
		\label{fig:baseline_exSGP_app}
	\end{centering}	
	\footnotesize Note. Standard errors in parenthesis are clustered at county-pair level.  ***, ** and * denote signifinance at the 1, 5 and 10 percent levels. Constants and the coefficients for the rest of the world are omitted.
\end{table}	

\begin{table}[ht!]
	\begin{centering}
		{\includegraphics[width=0.7\textwidth]{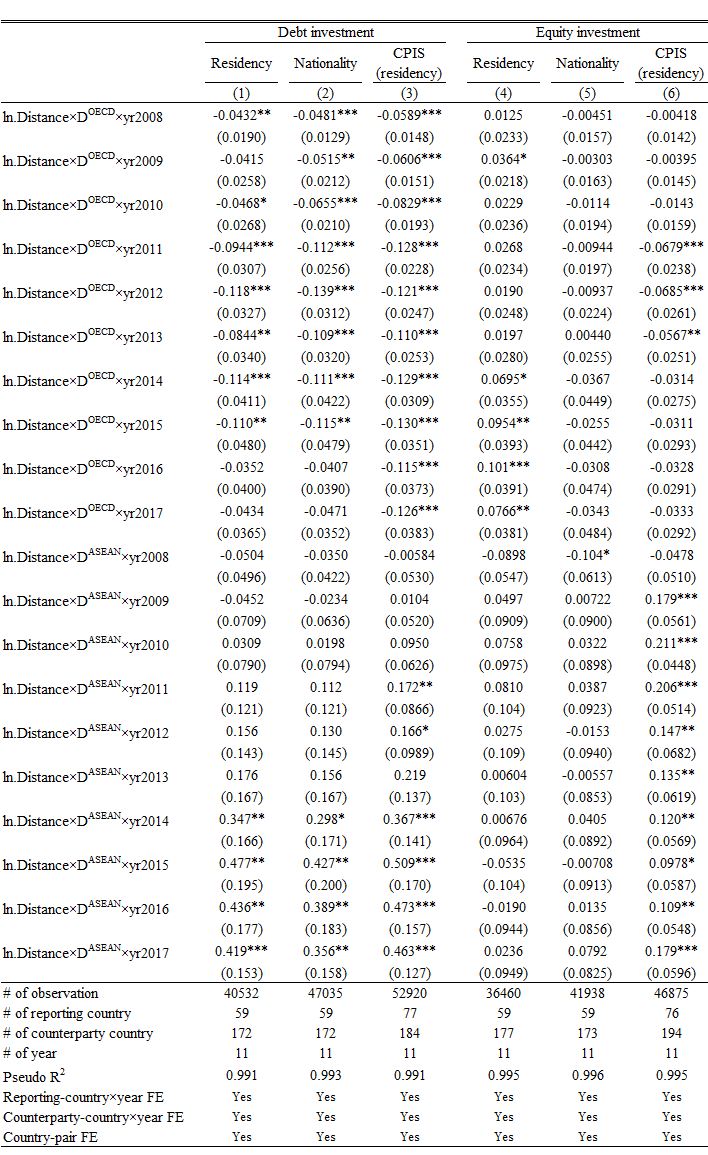}}
		\caption{Dynamic pattern of the coefficients}
		\label{fig:timeseries_app}
	\end{centering}	
	\footnotesize Note. Standard errors in parenthesis are clustered at county-pair level.  ***, ** and * denote signifinance at the 1, 5 and 10 percent levels. Constants and the coefficients for the rest of the world are omitted.
\end{table}	

\begin{table}[ht!]
	\begin{centering}
		{\includegraphics[width=0.7\textwidth]{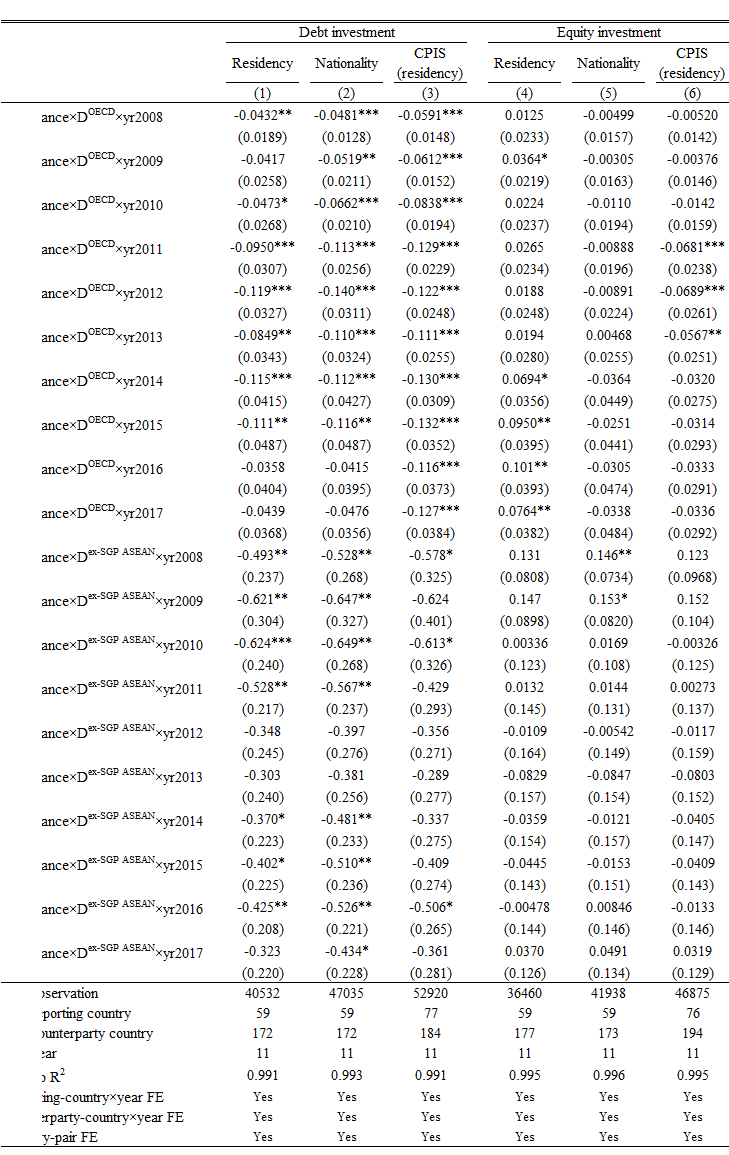}}
		\caption{Dynamic pattern of the coefficients (ASEAN ex-SGP dummy)}
		\label{fig:timeseries_exSGP_app}
	\end{centering}	
	\footnotesize Note. Standard errors in parenthesis are clustered at county-pair level.  ***, ** and * denote signifinance at the 1, 5 and 10 percent levels. Constants and the coefficients for the rest of the world are omitted.
\end{table}	

\clearpage

\bibliographystyle{ecta}

\bibliography{asean} 

\end{document}